\documentclass{aa}
\usepackage{natbib}
\usepackage{graphicx}
\usepackage{txfonts}

\usepackage{amsmath}	
\usepackage{amssymb}	
\usepackage{verbatim}
\usepackage{sansmath}
\usepackage{bm}

\newcommand{\de}{\text{d}}
\newcommand{\Msun}{M_{\odot}}
\newcommand{\kmsec}{\text{km}\text{s}^{-1}}

\newcommand{\pc}{\text{pc}} 
\newcommand{\Msunppcc}{\Msun\pc^{-3}}
\newcommand{\Msunppcsquare}{\Msun\pc^{-2}}

\newcommand{\mas}{\text{mas}}
\newcommand{\arcs}{\text{arcs}}
\newcommand{\masyr}{\text{mas}~\text{yr}^{-1}}

\newcommand{\popp}{\boldsymbol{\Psi}}
\newcommand{\objp}{\boldsymbol{\psi}}

\newcommand{\data}{\boldsymbol{d}}

\begin{document}

   \title{Measuring the local matter density using Gaia DR2}

   \author{A. Widmark
          \inst{1}
          }

   \institute{The Oskar Klein Centre for Cosmoparticle Physics, Department of
Physics, Stockholm University, AlbaNova, 10691 Stockholm, Sweden\\
              \email{axel.widmark@fysik.su.se}
             }

   \date{Received Month XX, 2018; accepted Month XX, XXXX}

 
  \abstract
   {}
   {We determine the total dynamical matter density in the solar neighbourhood using the second \emph{Gaia} data release (DR2).}
   {The dynamical matter density distribution is inferred in a framework of a Bayesian hierarchical model, which accounts for position and velocity of all individual stars, as well as the full error covariance matrix of astrometric observables, in a joint fit of the vertical velocity distribution and stellar number density distribution. This was done for eight separate data samples, with different cuts in observed absolute magnitude, each containing about 25,000 stars. The model for the total matter density does not rely on any underlying baryonic model, although we assumed that it is symmetrical, smooth, and monotonically decreasing with distance from the mid-plane.}
   {We infer a density distribution which is strongly peaked in the region close to the Galactic plane ($\lesssim 60~\pc$), for all eight stellar samples. Assuming a baryonic model and a dark matter halo of constant density, this corresponds to a surplus surface density of approximately 5--9 $\Msunppcsquare$. For the Sun's position and vertical velocity with respect to the Galactic plane, we infer $Z_\odot = 4.76\pm 2.27~\pc$ and $W_\odot = 7.24\pm 0.19~\kmsec$.}
   {These results suggest a surplus of matter close to the Galactic plane, possibly explained by an underestimated density of cold gas. We discuss possible systematic effects that could bias our result, for example unmodelled non-equilibrium effects, and how to account for such effects in future extensions of this work.}


   \maketitle
%

\section{Introduction}\label{sec:intro}

The distribution of matter in the Galactic disc has implications for the composition, dynamics and formation history of the Milky Way, as well as indirect and direct dark matter detection experiments \citep{Jungman:1995df,0004-637X-703-2-2275,2014MNRAS.444..515R}. The local matter distribution can be estimated dynamically by measuring the velocity and stellar number density distributions of the Galactic disc; assuming statistical equilibrium, these distributions are interrelated via the gravitational potential. Such measurements have been developed and applied by \cite{Kapteyn1922,Oort1932,1984ApJ...276..169B,1984ApJ...287..926B,KuijkenGilmore1989a,KuijkenGilmore1989b,KuijkenGilmore1989c,KuijkenGilmore1991,Creze1998,HolmbergFlynn2000,Bienayme:2005py,doi:10.1111/j.1365-2966.2012.21608.x,0004-637X-756-1-89,0004-637X-772-2-108,WidmarkMonari}. A history of pre-\emph{Gaia} dynamical mass measurements of the Galactic disc can be found in a review by \cite{Read2014}.

The distribution of matter in the solar neighbourhood is dominated by baryons, which are believed to be constituted by roughly equal densities of stars and gas \citep{Flynn:2006tm,2015ApJ...814...13M,0004-637X-829-2-126}. Out of all baryonic components, the density of cold gas (atomic and molecular) is the most uncertain. Precise constraints to the distribution of the total dynamical matter density can help constrain the distribution of baryonic components, supplementing other observations. Dynamical matter density measurements can also constrain or give insights into the particle nature of dark matter. \cite{Fan:2013tia,Fan:2013yva} have proposed what they call `double-disc dark matter': a dark matter sub-component with dissipative self-interaction, that can cool to form a thin dark disc, embedded inside a galaxy's stellar disc. The presence of such a thin dark disc in the Milky Way has been constrained in previous studies \citep{Kramer:2016dqu,Caputo:2017zqh,Schutz:2017tfp,Buch:2018qdr}, where the most stringent upper bound to its local surface density is approximately $7~\Msunppcsquare$ for a thin dark disc with a scale height of 30 pc (95~\% confidence).

With the quality of the second \emph{Gaia} data release (DR2), released in April 2018 \citep{2016A&A...595A...1G,2018A&A...616A...1G}, it is possible to fit more complicated models with greater precision than ever before. The depth and astrometric precision is significantly better than for previous data sets, and \emph{Gaia} DR2 also includes radial velocity measurements for 7.2 million stars. In this work we utilised the full potential of \emph{Gaia} DR2, in order to infer the dynamical matter distribution; we considered both position and velocity of all stars (not only those close to the Galactic plane), in a joint fit of the velocity and stellar number density distributions, while accounting for the full error covariance matrix of astrometric observables for each individual star. We did so for eight separate samples of stars within a 200 pc distance limited volume. This improves on previous works in terms of statistical modelling and sample size. Another key difference is that the total matter density inferred in this work does not rely on an underlying model for the baryonic and dark matter density distributions.

This article is structured as follows. In Sect.~\ref{sec:galacticmodel}, we outline our model of the solar neighbourhood stellar phase space density, and the distribution of baryonic matter. In Sect.~\ref{sec:data}, we present the stellar parameters that describe the stellar properties and the \emph{Gaia} data that constrains these properties, as well as our stellar sample construction and its resulting selection effects. We present our statistical model in Sect.~\ref{sec:statisticalmodel}, and our results in Sect.~\ref{sec:res}. Finally, in Sect.~\ref{sec:conclusion}, we discuss and conclude.

\section{Galactic model}\label{sec:galacticmodel}

In this section we describe our model of the phase-space distribution of stars in the solar neighbourhood. We used a system of Cartesian coordinates, where $\boldsymbol{x} = (x,y,z)$ denotes a position with respect to the Sun, where positive $x$ is in the direction of the Galactic centre, positive $y$ is in the Galaxy's rotational direction, and positive $z$ is in the direction of the Galactic north pole. The directions parallel and perpendicular to the Galactic plane are also referred to as horizontal and vertical directions. The Cartesian coordinate first order time derivative, $\boldsymbol{v} \equiv (u,v,w) \equiv (\dot{x},\dot{y},\dot{z})$, is velocity in the Sun's rest frame. When writing $Z$, we refer to height with respect to the Galactic plane. Similarly, when writing $W$, we refer to the vertical velocity in the Galactic rest frame.

In this work, we are volume limited to 200 pc in distance from the Sun, meaning that we also stay very close to the Galactic plane. Within this small volume, we made the following assumptions: the stellar phase-space density $f$ is invariant under translation in $x$ and $y$; the gravitational potential is separable in the horizontal and vertical directions, and the vertical energy is an integral of motion \citep{BT2008}; stars are in steady-state, such that their distribution in space and vertical velocity is completely described by the gravitational potential and their distribution in vertical energy.

Any coupling between the vertical and horizontal velocity distributions are thus neglected; in Jeans theory such couplings are known as the `axial' and `tilt' terms, which are believed to be small this close to the Galactic plane \citep{budenbender,Sivertsson:2017rkp}. \cite{Garbari2011} has a useful discussion on the validity of these assumptions and test them on simulations. They find that systematic errors become significant at distances of about 500 pc above the galactic plane. Hence, the problem of inferring the gravitational potential is reduced to one dimension and can be formulated in terms of $f_\perp(W,Z)$, which is the vertical velocity distribution as a function of height with respect to the Galactic plane. This distribution is discussed in detail below.

{\renewcommand{\arraystretch}{1.6}
\begin{table*}[ht]
	\centering
	\caption{Population parameters, the object parameters and data of each stellar object, the parameters and distributions with implicit dependence on the population parameters, and the fixed background distributions of our model.}
	\label{tab:parameters}
    \begin{tabular}{| l | l |}
		\hline
		$\popp$  & Population parameters (11 degrees of freedom) \\
		\hline
		$\rho_{\{1,2,3,4\}}$ & mid-plane densities of matter components with different scale heights \\
		$c_{j=\{2,3\}}$ & two weights of the vertical velocity distribution Gaussians, also setting $c_1 = 1-c_2-c_3$ \\
		$\sigma_{j=\{1,2,3\}}$ & three Gaussian widths of the vertical velocity distribution \\
		$Z_\odot$ & height of the Sun above the Galactic plane \\
		$W_\odot$ & vertical velocity of the Sun in the Galactic rest frame  \\
        \hline
        \hline
        $\objp_{i=1,...,N}$  & Stellar parameters (7 degrees of freedom per stellar object) \\
        \hline
        $\mathbf{x} = (x,y,z)$ & spatial position in the solar rest frame  \\
        $\mathbf{v} = (u,v,w)$ & velocity in the solar rest frame \\
        $M_G$ & absolute magnitude in the \emph{Gaia} $G$-band  \\
        \hline
        \hline
        $\data_{i=1,...,N}$ & Data \\
        \hline
        $m_G$ & apparent magnitude in the \emph{Gaia} $G$-band \\
        $l$, $b$ & Galactic longitude and latitude \\
        $\hat{\varpi}$ & observed parallax \\
        $\hat{\mu}_l$, $\hat{\mu}_b$  & observed proper motions \\
        $\hat{\boldsymbol{\Sigma}}$ & error covariance matrix for proper motions and parallax \\
        $\hat{v}_{RV}$, $\hat{\sigma}_{RV}$  & observed radial velocity and its uncertainty (available only for a subset of stars) \\
        \hline
        \hline
        & Parameters and distributions that depend on the population parameters \\
        \hline
        $\Phi(Z)$ & gravitational potential \\
        $f_\perp(W,Z)$ & vertical velocity distribution, which integrates to $n(Z)$ \\
        $n(Z)$ & relative stellar number density \\
        $Z = z+Z_\odot$ & height with respect to the Galactic plane \\
        $W=w+W_\odot$ & vertical velocity in the Galactic rest frame \\
		\hline
        \hline
        & Fixed background distributions \\
        \hline
        $F(M_G)$ & luminosity function \\
        $f_\parallel(u,v)$ & velocity distribution parallel to the Galactic plane, parametrised by $a_k$, $\boldsymbol{g}_k$, and $\boldsymbol{G}_k$ \\
        $C(l,b,m_G)$ & DR2 completeness \\
        $h(\hat{\sigma}_\varpi\, |\, m_G,b)$ & density of parallax uncertainties, as a function of apparent magnitude and Galactic latitude \\
        $\Xi(\mathbf{x})$ & $G$-band extinction due to dust \\
        \hline
	\end{tabular}
\end{table*}}

Vibrational modes of the Galactic disc are probably not of importance this close to the Galactic plane. \cite{breathingmode} has shown that a breathing mode can affect the inference of the dynamical matter density at kpc heights, but this is at significantly greater distances than what is relevant for this work.

With \emph{Gaia} DR2, many studies demonstrate phase-space substructure both in the local region and in the larger Galactic structure, with moving groups that exhibit incomplete vertical phase-mixing \citep{gaia_kinematics,hercules_group1,hercules_group2,coma_berenices,riding_substructure,open_clusters,wrinkles,DM_substructure,BovyAssym}. It is difficult to quantify the significance and potential bias that arises from the assumption of statistical equilibrium. In Sect.~\ref{sec:conclusion}, we discuss possible ways to relax them in future extensions of this work.

\subsection{Vertical velocity and stellar number density}

The population parameters of our model parametrise the vertical velocity distribution and how it changes with height with respect to the Galactic plane, as well as how the stellar number density varies with height. There are four parameters that parametrise the matter density as a function of height above the Galactic plane, $\rho_{\{1,2,3,4\}}$; five parameters that describe the distribution of vertical velocity in the plane, $c_2$, $c_3$, and $\sigma_{\{1,2,3\}}$; one parameter is the vertical height of Sun with respect to the Galactic plane, $Z_\odot$; one parameter is the vertical velocity of the Sun in the Galactic rest frame, $W_\odot$. The eleven population parameters are listed in Table~\ref{tab:parameters}.

The gravitational potential $\Phi(Z)$ is given by the total dynamical matter density of the Galactic disc, according to the one-dimensional Poisson equation
\begin{equation}\label{eq:Poisson}
	\frac{\partial^2\Phi(Z)}{\partial Z^2} = 4\pi G \rho(Z),
\end{equation}
which is valid close to the plane. The gravitational potential is defined to be zero and have a zero valued first order derivative in the Galactic plane ($Z=0$). The matter density as a function of height is taken as a sum of four matter components with different scale heights, according to
\begin{equation}
\begin{split}
    \rho(Z) =
    & \; \rho_1\, \text{sech}^2 \left( \frac{Z}{30~\pc} \right)+
    \rho_2\, \text{sech}^2 \left( \frac{Z}{60~\pc} \right) \\
    & + \rho_3\, \text{sech}^2 \left( \frac{Z}{120~\pc} \right)+
    \rho_4,
\end{split}
\end{equation}
where these densities are in units of mass over volume.

The vertical velocity distribution in the Galactic mid-plane is described by population parameters $c_2$, $c_3$, and $\sigma_{\{1,2,3\}}$, according to
\begin{equation}
	f_\perp(W,Z=0) = \sum_{j=1}^3 c_j\frac{\exp\left(-\dfrac{W^2}{2\sigma_j^2}\right)}{\sqrt{2\pi\sigma_j^2}},
\end{equation}
where $c_1 \equiv 1-c_2-c_3$, with constraints $\sum c_j =1$ and $c_j\geq 0$, $\forall j$. This function integrates to unity. The index $j$ is only ever used to label the Gaussian mixture of the vertical velocity distribution.

Leaving the plane, the vertical velocity distribution becomes
\begin{equation}
	f_\perp(W,Z) = \sum_{j=1}^3 c_j\frac{\exp\left[-\dfrac{W^2+2\Phi(Z)}{2\sigma_j^2}\right]}{\sqrt{2\pi\sigma_j^2}}.
\end{equation}
When $Z\neq 0$, integrating $f_\perp(Z,W)$ over $W$ does not give unity, but is equal to the relative change to the stellar number density,
\begin{equation}\label{eq:nuofz}
 	n(Z)=\sum_{j=1}^3 c_j \exp\left[-\frac{\Phi(Z)}{\sigma_j^2}\right].
\end{equation}

\subsection{Horizontal velocity distribution}

The stellar velocities parallel to the Galactic plane, written $u$ and $v$ in the solar rest frame, are not of primary importance for this study, although necessary to account for if we want to use the velocity information of all stellar objects. For stellar objects that are missing radial velocity measurements, the horizontal velocities must be marginalised over. In order to perform this marginalisation in a computationally efficient way, we made the assumption that the vertical and horizontal velocity distributions are separable, such that
\begin{equation}\label{eq:separable_velocities}
	f(\boldsymbol{x},\boldsymbol{v}) \propto
    f_\parallel(u,v) f_\perp(W,Z).
\end{equation}
This approximation is a simplification and not actually true (even though the gravitational potential is separable in the sample volume); stars with high vertical energies also have a larger velocity dispersion in the horizontal directions, and the vertical energy is also correlated with the mean azimutal velocity ($v$). Such correlations are not accounted for in our model, but this is expected to have a minor effect. For the range of apparent magnitudes used in this work, at least 80 per cent of stars have radial velocity measurements, meaning that full three-dimensional velocity information is available. For most stellar objects, the velocity uncertainty in any direction is significantly smaller than $1~\kmsec$. For such objects, the horizontal velocity distribution has a negligible impact; it mainly acts as a prior to the objects that are missing radial velocity measurements. For this reason, the approximation of separability, as expressed in Eq.~\eqref{eq:separable_velocities}, is not expected to produce a significant bias in our result. We also quantified this bias, by performing a fit where we only consider the velocity information of stars with good radial velocity measurements ($\hat{\sigma}_{RV}<3~\kmsec$), and disregard the horizontal velocity distribution completely. This is described in more detail in Sect.~\ref{sec:res}.

The horizontal velocity distribution is not well described by a separable bivariate Gaussian; instead, we fitted a joint $(u,v)$ distribution with a Gaussian mixture model, written $f_\parallel(u,v)$. This Gaussian mixture model is parametrised by five weights, $a_k$; five two-dimensional mean values, $\boldsymbol{g}_k = \{g_{u,k},g_{v,k}\}$; and five $2\times 2$ covariance matrices, $\boldsymbol{G}_k$. The index $k$ is only ever used to label the Gaussian mixture of this velocity distribution. Because $f_\parallel(u,v)$ is formulated in terms of velocities in the Sun's rest frame, it accounts for the Sun's velocity parallel to the plane. Having the velocity distribution on a Gaussian form is useful computationally, as the integration over three-dimensional velocity can be done analytically. Details of how the Gaussian mixture model is fitted is found in Appendix~\ref{app:uvGMM}.

\begin{figure*}
	\centering
	\includegraphics[width=1.\linewidth]{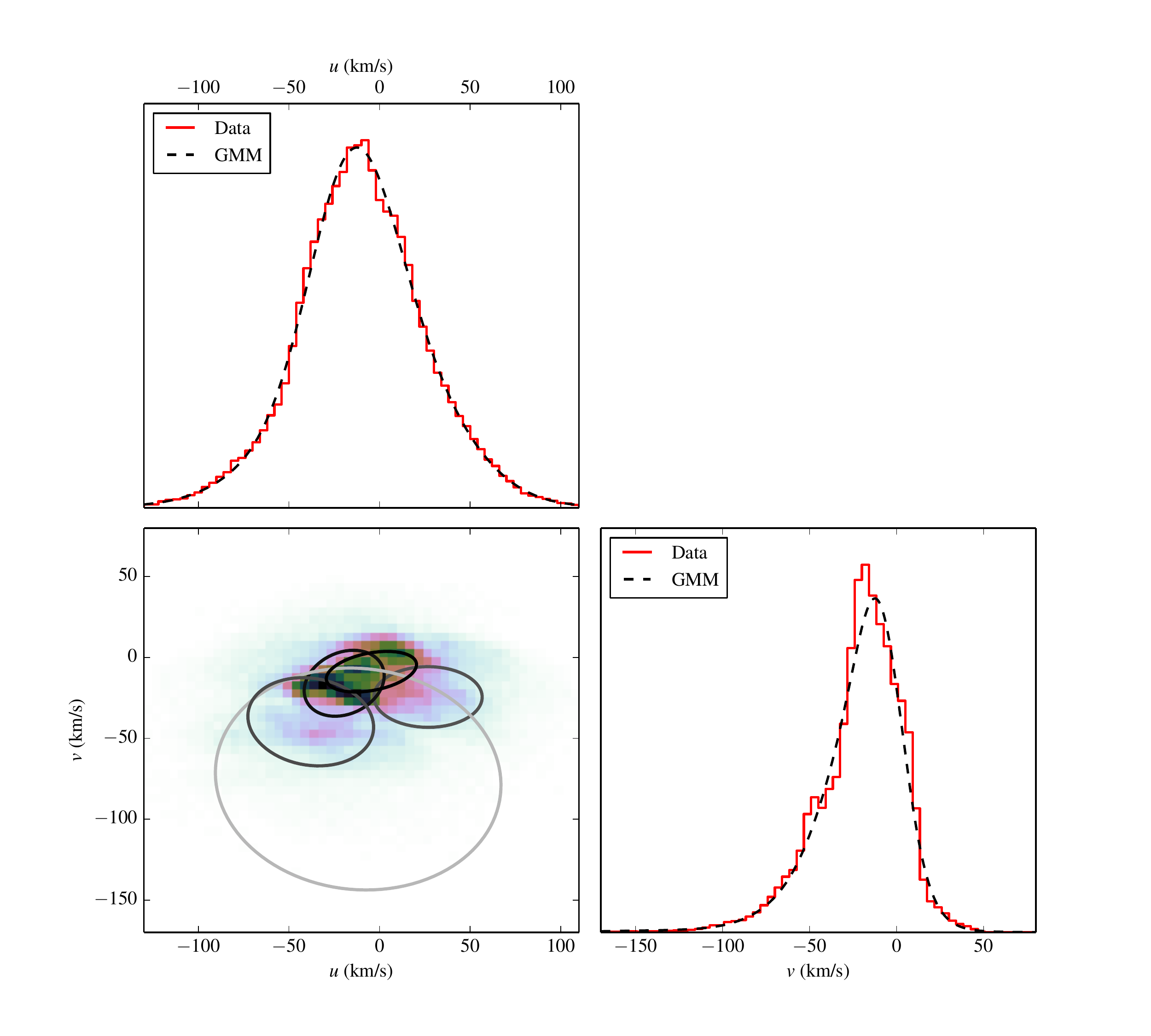}
    \caption{Gaussian mixture model and data over $u$ and $v$ velocities. In the bottom-left panel, the data is plotted as a 2D histogram, overlaid by the 1-sigma ellipses of the five Gaussians of the mixture model. The shading of the ellipses correspond to the weight of the respective Gaussians, where a darker colour correspond to a larger weight. In the top and bottom-right panels, the fitted model and the data are marginalised to 1D histograms over $u$ and $v$.}
    \label{fig:uvGMM}
\end{figure*}

The Gaussian mixture model is shown in Fig.~\ref{fig:uvGMM}, in two-dimensional and projected one-dimensional histograms over $u$ and $v$. The overall structure of the joint velocity distribution is well described by the Gaussian mixture model, although there are some substructures that are not captured, most clearly seen in the projection over $v$. We expect these smaller inaccuracies to have a negligible effect on our end result, much smaller than other systematic uncertainties.

\subsection{Dust}\label{sec:dust}

Extinction due to dust is most likely a negligible effect in this analysis. To a distance of at least 100 pc, the solar neighbourhood is almost devoid of dust. Dust reddening is typically $\text{E}(B-V)\simeq 0.01$ at 100 pc distance, and $\text{E}(B-V)\simeq 0.025$ at 200 pc, with some smaller patches of more severe reddening \citep{2017A&A...606A..65C,2018MNRAS.478..651G}. We use a conversion factor of 3.1 for going from $\text{E}(B-V)$ reddening to \emph{Gaia} $G$-band extinction \citep{dust_coeff,2018A&A...614A..19D}. Although the extinction factor has some dependence on colour, such effects are negligible compared to other dust uncertainties.

The dust extinction is significantly smaller than the absolute magnitude sample ranges. Despite being having a minor effect, the three-dimensional dust map of \cite{2017A&A...606A..65C} was used in this analysis, although uncertainties to the dust distribution were ignored. The resulting $G$-band extinction is a function of three-dimensional spatial position, and is written $\Xi(\boldsymbol{x})$.

\subsection{The luminosity function $F(M_G)$}

\begin{figure}
	\centering
	\includegraphics[width=.8\columnwidth]{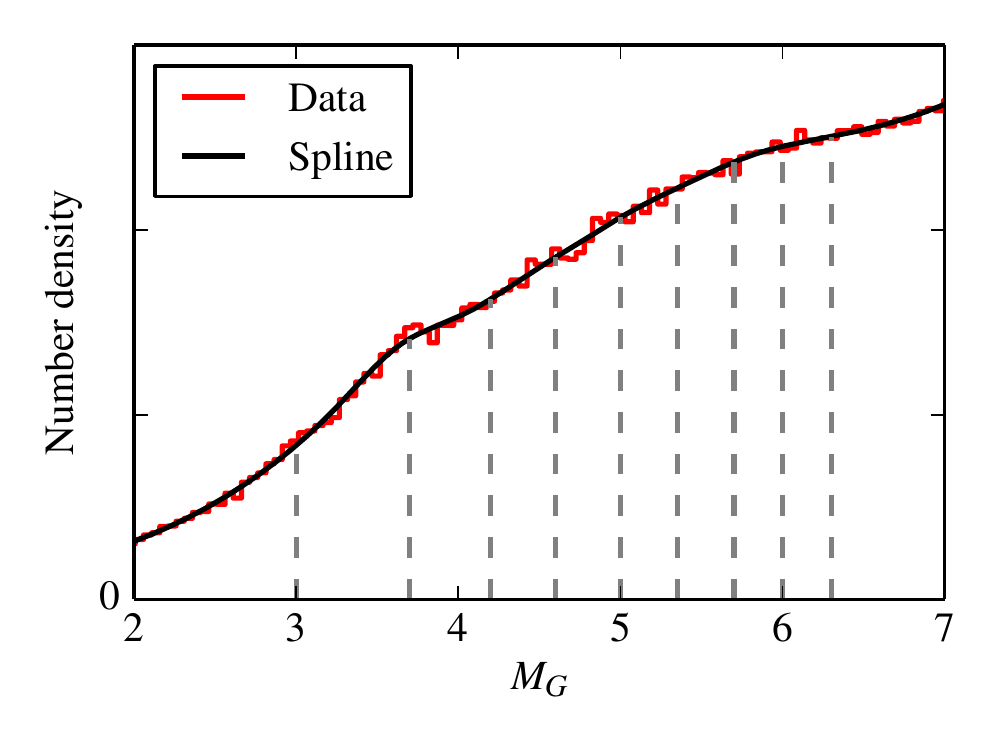}
    \caption{Luminosity function $F(M_G)$, describing the number density of stars in the solar neighbourhood, as a function of absolute magnitude in the \emph{Gaia} $G$-band. The solid lines are the binned data and the spline function $F(M_G)$. The dashed lines correspond to the absolute magnitude bounds by which the four data samples are constructed. The function is not normalised and need not be.}
    \label{fig:F(M)}
\end{figure}

The luminosity function describes the number density of stars as a function of absolute magnitude. We fitted this distribution to data and treated it as a fixed background distribution. The data consists of stars with a distance in range 100--200 pc, binned in terms of $\hat{M}_G$ (calculated from the observed apparent magnitude, observed position, and dust map). The function $F(M_G)$ is a spline on this binned data, as is shown in Fig.~\ref{fig:F(M)}.

The luminosity function is important for calculating the normalisation factor $\bar{N}$, which is discussed in Sect.~\ref{sec:norm}. It also gives some power of inference, albeit very weak, in terms of the true distance of a star, as brighter stars are less numerous than dimmer stars.

\subsection{Baryonic density}\label{sec:baryons}

In this work, we infer the local dynamical density without relying on any underlying model for the baryonic density. However, a baryonic density model is presented in this section, in order to be able to compare the inferred dynamical density with the expected distribution of matter. The baryonic density model was taken from \cite{Schutz:2017tfp}, who have compiled a table of baryonic components using results from \cite{Flynn:2006tm,2015ApJ...814...13M,0004-637X-829-2-126}. A list of the baryonic density components is presented in Table~\ref{tab:baryonic_components}.

Summing the densities of these components (and their uncertainties in quadrature), the total baryonic mid-plane density is equal to $0.0889\pm 0.0071~\Msunppcc$. The total baryonic density, as a function of height over the Galactic plane, is equal to
\begin{equation}
    \rho_b = \sum_{t}
    \rho_{t}\exp\left[ -\frac{\Phi(Z)}{\sigma_{t}^2} \right],
\end{equation}
where the index $t$ denotes the respective baryonic components.

\begin{figure}
	\centering
	\includegraphics[width=\columnwidth]{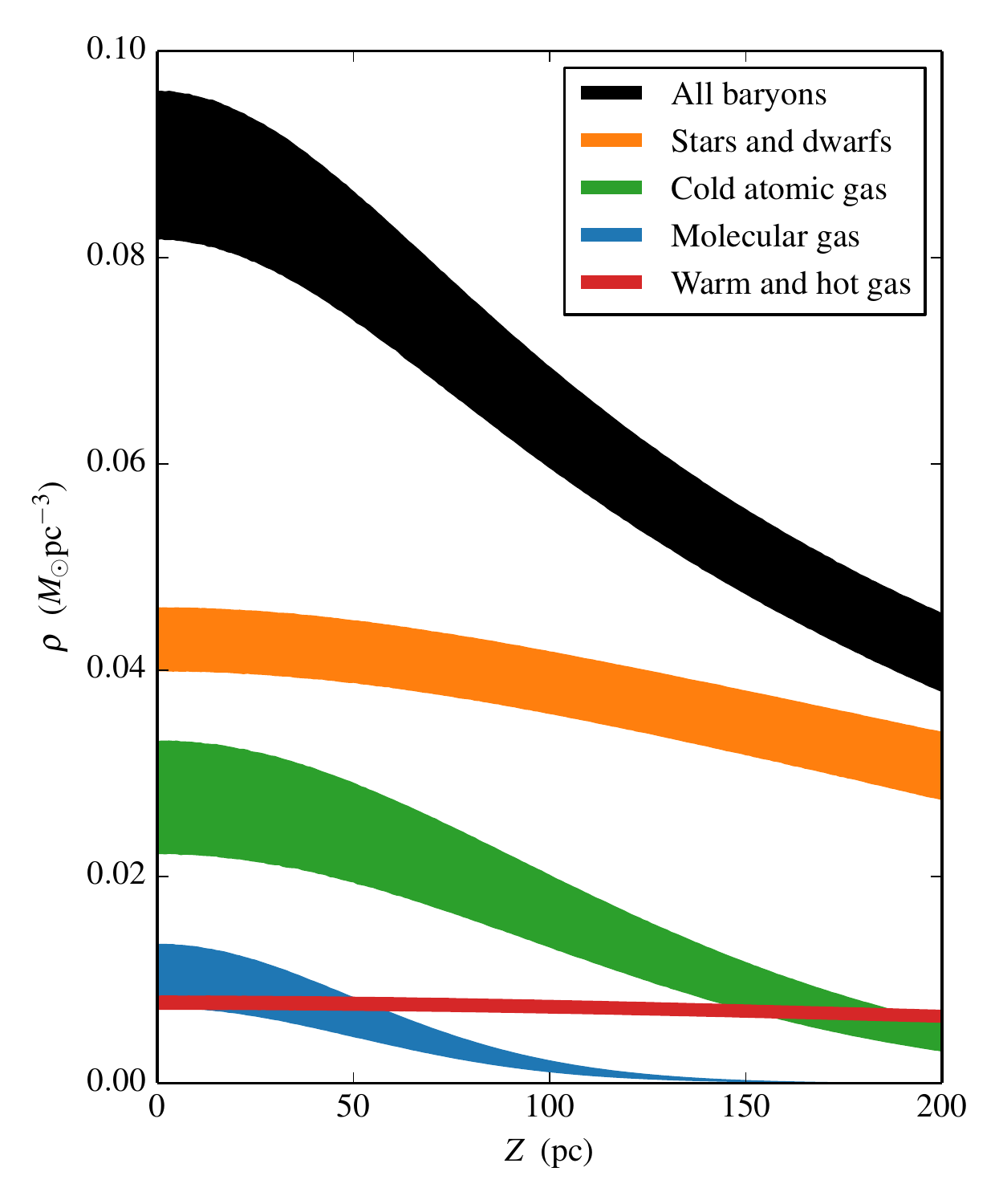}
    \caption{Baryonic matter density, as a function of height above the Galactic plane. Five coloured 1-$\sigma$ bands show the total baryonic density, as well as a division into four separate groups: the `stars and dwarfs' include all types of stars, white dwarfs, and brown dwarfs; the `cold atomic gas' and `molecular gas' correspond directly to their respective single components of Table~\ref{tab:baryonic_components}; the `warm and hot gas' correspond to the warm atomic gas and hot ionised gas components.}
    \label{fig:baryonic_density}
\end{figure}

The baryonic density distribution is shown in Fig.~\ref{fig:baryonic_density}, as a sum total and also split into different components. The molecular gas and cold atomic gas components are the coldest components, hence they have the smallest scale heights, and their respective mid-plane densities are more uncertain than that of any other baryonic component. The bands represent 1-$\sigma$ error bars, which accounts for uncertainties of both the baryonic component mid-plane densities as well as the baryonic component velocity dispersions. In this model, the gravitational potential is set by a homogeneous dark matter halo density of $0.01~\Msunppcc$ and the baryonic density itself. If the gravitational well is steeper, due to a higher mid-plane density for example, the baryonic density will fall off quicker with distance from the Galactic plane \citep[referred to as `pinching' by][]{0004-637X-829-2-126}.

{\renewcommand{\arraystretch}{1.6}
\begin{table}
\caption{Mid-plane densities $\rho_{t}$ and velocity dispersions $\sigma_{t}$ of baryonic matter components.}
\label{tab:baryonic_components}      
\centering          
\begin{tabular}{l l l} 
\hline
Component & $\rho_{t}~(\Msunppcc)$ & $\sigma_{t}~(\kmsec)$ \\
\hline
Molecular gas & $0.0104\pm 0.00312$ & $3.7\pm 0.2$ \\
Cold atomic gas & $0.0277\pm 0.00554$ & $7.1\pm 0.5$ \\
Warm atomic gas & $0.0073\pm 0.0007$ & $22.1\pm 2.4$ \\
Hot ionised gas & $0.0005\pm 0.00003$ & $39.0\pm 4.0$ \\
Giant stars & $0.0006\pm 0.00006$ & $15.5\pm 1.6$ \\
Stars, $M_V<3$ & $0.0018\pm 0.00018$ & $7.5\pm 2.0$ \\
Stars, $3<M_V<4$ & $0.0018\pm 0.00018$ & $12.0\pm 2.4$ \\
Stars, $4<M_V<5$ & $0.0029\pm 0.00029$ & $18.0\pm 1.8$ \\
Stars, $5<M_V<8$ & $0.0072\pm 0.00072$ & $18.5\pm 1.9$ \\
Stars, $M_V>8$ & $0.0216\pm 0.0028$ & $18.5\pm 4.0$ \\
White dwarfs & $0.0056\pm 0.0010$ & $20.0\pm 5.0$ \\
Brown dwarfs & $0.0015\pm 0.0005$ & $20.0\pm 5.0$ \\
\hline                  
\end{tabular}
\end{table}}

There are improvements that can be made to this model. The list of stellar components is not based on \emph{Gaia} data, but less precise surveys. Furthermore, the vertical velocity distribution of the different stellar populations are parametrised by a single velocity dispersion; it is clear from the \emph{Gaia} data that the vertical velocity distribution is not well described by a single Gaussian. This is important to account for, especially further from the Galactic plane, in order to correctly model the density of stars as a function of height. In this work, we stay close to the Galactic plane, which is why uncertainties to the molecular and atomic gas densities are more critical.

\section{Stellar parameters, data and selection}\label{sec:data}

In our model, each stellar object has seven degrees of freedom: three-dimensional position, $\boldsymbol{x}$; three-dimensional velocity, $\boldsymbol{v}$; and absolute magnitude in the \emph{Gaia} $G$-band, $M_G$. The properties of the $i$th stellar object are encapsulated in $\objp_i$, where $i$ ranges from one to $N$, where $N$ is the total number of stellar objects in a sample. We use the term `stellar object' rather than `star', as some of them are unresolved multiple stellar systems.

The \emph{Gaia} observables, written with hats, that constrain the stellar parameters are: apparent magnitude, $\hat{m}_G$; Galactic longitude and latitude, $\hat{l}$ and $\hat{b}$; observed parallax, $\hat{\varpi}$; observed proper motions in the latitudinal and longitudinal directions, $\hat{\mu}_l$ and $\hat{\mu}_b$; and observed radial velocity, $\hat{v}_{RV}$. The radial velocity is available only for a subset of stars in \emph{Gaia}. The data of a stellar object is written $\data_i$, where $i$ ranges from one to $N$, just like for the stellar parameters.

The observational uncertainties on apparent magnitude and angular coordinates are small and can be neglected. The former has uncertainties of around 1 mmag, with possible systematic errors up to 10 mmag \citep{photo_systematics}.

There are systematic errors to the astrometric measurements of the \emph{Gaia} DR2 catalogue. According to \cite{astrometry_systematics}, such systematic errors are smaller than $0.1~\mas$ for parallax measurements and $0.1~\masyr$ for proper motion measurements. In addition to a global parallax zero-point shift of $-0.03~\mas$, which we corrected for, there are spatially correlated systematic errors as large as $0.04~\mas$ for parallax and $0.07~\masyr$ for proper motion, on angular scales as large as roughly 20 degrees. Because we are distance limited to 200 pc, parallax errors of order $0.1~\mas$ correspond to relative errors in distance of at most a few per cent. Systematic proper motion errors as large as $0.1~\masyr$ correspond to velocity errors $\lesssim 0.1~\kmsec$. The radial velocity measurements of \emph{Gaia} have been compared with several ground-based catalogues, as detailed in \cite{rv_systematics}; the difference in radial velocity median values with respect to these catalogues is of the order $0.1~\kmsec$. Even assuming a worst case scenario of very large systematic errors, such errors are not large enough to significantly bias the result of this work.

The uncertainties of the proper motions and parallax measurements are encoded in an error covariance matrix
\begin{equation}
	\hat{\boldsymbol{\Sigma}} = 
    \begin{bmatrix}
	\hat{\sigma}_{\mu_l}^2 & \hat{\rho}_{\mu_l\mu_b}\hat{\sigma}_{\mu_l}\hat{\sigma}_{\mu_b} & \hat{\rho}_{\mu_l\varpi}\hat{\sigma}_{\mu_l}\hat{\sigma}_{\varpi} \\
    \hat{\rho}_{\mu_l\mu_b}\hat{\sigma}_{\mu_l}\hat{\sigma}_{\mu_b} & \hat{\sigma}_{\mu_b}^2 & \hat{\rho}_{\mu_b\varpi}\hat{\sigma}_{\mu_b}\hat{\sigma}_{\varpi} \\
    \hat{\rho}_{\mu_l\varpi}\hat{\sigma}_{\mu_l}\hat{\sigma}_{\varpi} & \hat{\rho}_{\mu_b\varpi}\hat{\sigma}_{\mu_b}\hat{\sigma}_{\varpi} & \hat{\sigma}_{\varpi}^2
   \end{bmatrix},
\end{equation}
where $\hat{\sigma}_{}$ is an uncertainty, and $\hat{\rho}_{}$ is a correlation coefficient. The uncertainty on the radial velocity is written $\hat{\sigma}_{RV}$, which is uncorrelated with other observables.

The likelihood of observing the data $\data_i$, given a stellar object with stellar parameters $\objp_i$, is interpreted as being proportional to
\begin{equation}
\begin{split}
	& \text{Pr}(\data_i \, | \objp_i) \propto \\
    & \delta(m_G-\hat{m}_G)\,\delta(l-\hat{l})\,\delta(b-\hat{b})\,\mathcal{N}\left(
    \begin{bmatrix}
    	\mu_l - \hat{\mu}_l \\
        \mu_b - \hat{\mu}_b \\
        \varpi - \hat{\varpi} \\
    \end{bmatrix}, \hat{\boldsymbol{\Sigma}}
    \right)\,
    \mathcal{G}(\hat{v}_{RV},\hat{\sigma}_{RV}),
\end{split}
\end{equation}
where $\delta(...)$ is the Dirac delta function, $\mathcal{N}(...)$ is the multivariate Gaussian distribution, and $\mathcal{G}(...)$ is the one-dimensional Gaussian distribution. Quantities without hats correspond to the observable's true value, as given by the stellar parameters; for example, the parallax $\varpi$ is given by the inverse of the true distance. The last factor of this expression is dropped when no radial velocity measurement is available. The multivariate Gaussian distribution is defined by
\begin{equation}\label{eq:multivariate_Gaussian}
	\mathcal{N}(\boldsymbol{p},\boldsymbol{\Sigma}_{\boldsymbol{p}}) \equiv
    \frac{\exp\left(-\dfrac{1}{2} \boldsymbol{p}^\top\boldsymbol{\Sigma}_{\boldsymbol{p}}^{-1}\boldsymbol{p} \right)}{\sqrt{(2\pi)^q \, | \boldsymbol{\Sigma}_{\boldsymbol{p}} |}},
\end{equation}
where $\boldsymbol{p}$ is a column vector of length $q$, and $\boldsymbol{\Sigma}_{\boldsymbol{p}}$ is a $q\times q$ matrix with determinant $| \boldsymbol{\Sigma}_{\boldsymbol{p}} |$. The one-dimensional Gaussian distribution is defined by
\begin{equation}\label{Gaussian}
	\mathcal{G}(p,\sigma_{p}) \equiv
    \frac{\exp\left(-\dfrac{p^2}{2\sigma_p^2} \right)}{\sqrt{2\pi  \sigma_{p}^2 }}.
\end{equation}

In the rest of this article, the hats on $\hat{m}_G$, $\hat{l}$, and $\hat{b}$ are dropped, to signify that their uncertainties were neglected. The stellar parameters and the data are listed in Table~\ref{tab:parameters}.

\subsection{Sample construction}
\label{sec:samplecuts}

In terms of constructing data samples, the general approach of this work was to make a minimal amount of cuts, directly on \emph{Gaia} observables, in order not to invoke any uncontrolled biases. Rather than cleaning the data of objects with large observational uncertainties, we carefully accounted for the effects of the overall uncertainty distribution.

We constructed eight separate stellar samples from \emph{Gaia} DR2, by making cuts in observed distance and observed absolute magnitude, according to
\begin{equation}
	\hat{d}_\text{min} <  \frac{\arcs \times \pc}{\hat{\varpi}} < \hat{d}_\text{max}
\end{equation}
and
\begin{equation}
	\hat{M}_{G,\text{min}} < m_G-5\Bigg[\log_{10}\Bigg(\frac{\arcs}{\hat{\varpi}}\Bigg)-1\Bigg]-\Xi(\hat{\boldsymbol{x}}) < \hat{M}_{G,\text{max}},
\end{equation}
where it is implicit that the observed position $\hat{\boldsymbol{x}}$ is given by $l$, $b$ and $\hat{\varpi}$. The limits in distance are $\hat{d}_\text{min}=100~\pc$ and $\hat{d}_\text{max}=200~\pc$ for all samples.

The dynamical assumptions made in Sect.~\ref{sec:galacticmodel} are expected to hold only close to the Galactic plane, which is why we set a limit of 200 pc in distance. Furthermore, dust extinction becomes more severe and more uncertain with greater distances (see Sect.~\ref{sec:dust}). We set a lower distance bound of 100 pc, such that the sample volume is the shape of a shell. The main reason for this is that the completeness is lower and more poorly modelled for bright objects. There can also be a number count bias due to multiple stellar systems. Treating unresolved binaries as single stellar objects is not in itself a problem, but a binary system of a certain orbital separation will be resolved only if it is sufficiently close. For this reason, not considering the volume within 100 pc distance reduces the number count bias that can arise from having a sample volume that spans a wide range of distances.

The limits in absolute magnitude for the respective samples are listed in Table~\ref{tab:samples}. Given these data cuts, the stellar objects will have an apparent magnitude in range 8--15. The ranges in absolute magnitude are chosen such that each sample has a roughly equal number of stellar objects. The range of absolute magnitude (3.0--6.3) corresponds to stars that are mainly of F and G classification.

{\renewcommand{\arraystretch}{1.6}
\begin{table}
\caption{List of stellar samples, with sample names, range of observed absolute magnitude, number of stellar objects $N$, the fraction of stellar objects with available radial velocity measurements, and the fraction of stellar objects with radial velocity uncertainties smaller than $3~\kmsec$}.         
\label{tab:samples}
\centering          
\begin{tabular}{l l l l l} 
\hline
Name & $\hat{M}_G$ & $N$ & RV \% & $\hat{\sigma}_{RV}<3~\kmsec$ \% \\
\hline
S1 & 3.0--3.7 & 27,010 & 81.2 & 75.7 \\
S2 & 3.7--4.2 & 25,812 & 85.1 & 80.2 \\
S3 & 4.2--4.6 & 23,895 & 88.5 & 83.8 \\
S4 & 4.6--5.0 & 26,847 & 90.5 & 85.8 \\
S5 & 5.0--5.35 & 25,568 & 91.2 & 86.3 \\
S6 & 5.35--5.7 & 27,591 & 90.3 & 84.7 \\
S7 & 5.7--6.0 & 24,730 & 89.9 & 83.3 \\
S8 & 6.3--6.3 & 25,531 & 88.5 & 80.5 \\
\hline                  
\end{tabular}
\end{table}}

\subsection{Selection function}
\label{sec:selection}

The selection function consists of the two parts: the data cuts of the sample construction and the completeness function. The completeness of \emph{Gaia} DR2, written $C(l,b,m_G)$, is a function of angular position and apparent magnitude. It was calculated from a cross-match with the 2MASS catalogue, made by Rybizki et. al.\footnote{\url{https://github.com/jan-rybizki/gdr2_completeness}} This selection function is very coarse in apparent $G$-band magnitude range, split in bins of 8--12 and 12--15 mag, where the latter has marginally better completeness. An error in the \emph{Gaia} completeness function could affect the result, probably in the sense of a bias that is shared between samples of similar brightness. However, the all-sky completeness is evaluated to around 99 per cent, which has a negligible effect. The selection function acts solely on the data, according to
\begin{equation}\label{eq:selectionfunction}
\begin{split}
	S(\data_i) & =
    C(l,b,m_G) \times
    \Theta\left( \frac{\arcs \times \pc}{\hat{\varpi}} > \hat{d}_\text{min}  \right)\,
    \Theta\left( \frac{\arcs \times \pc}{\hat{\varpi}} < \hat{d}_\text{max}  \right)\\
    & \times
    \Theta\left\{ m_G-5\Bigg[\log_{10}\Bigg(\frac{\arcs}{\hat{\varpi}}\Bigg)-1\Bigg]-\Xi(\hat{\boldsymbol{x}}) > \hat{M}_{G,\text{min}} \right\} \\
    & \times
    \Theta\left\{ m_G-5\Bigg[\log_{10}\Bigg(\frac{\arcs}{\hat{\varpi}}\Bigg)-1\Bigg]-\Xi(\hat{\boldsymbol{x}}) < \hat{M}_{G,\text{max}}  \right\},
   \end{split}
\end{equation}
where $\Theta(...)$ is the Heaviside step function.

\subsection{Parallax uncertainties}\label{sec:parallax_errors}

In order to calculate the expected number of stellar objects that are included in the sample ($\bar{N}$, discussed at length in Sect.~\ref{sec:norm}), it is necessary to model the distribution of observational uncertainties that affect selection. The one observational quantity with significant uncertainties that affects selection is the parallax.

The distribution of parallax uncertainties was modelled as a function of apparent magnitude in the \emph{Gaia} $G$-band and Galactic latitude $b$, and is written $h(\hat{\sigma}_\varpi \, | \, m_G,b)$. Using only a cut in distance with an upper limit of 200 pc, the \emph{Gaia} DR2 catalogue is split into bins of apparent magnitude of width 0.5 mag, and ten bins in Galactic latitude that divide the sky in regions of equal area. There is no need to also divide the data into bins in Galactic longitude, as this will be integrated out when calculating the $\bar{N}$.

The median parallax uncertainty in these bins is about 0.05 mag, but there is a tail of higher uncertainties; the 95th percentile of these bins vary between 0.2--0.7 mag. This tail is especially pronounced for dimmer stars close to the Galactic plane, presumably due to stellar crowding.

\section{Statistical model}\label{sec:statisticalmodel}

In this section we present the statistical model and sampling technique by which we inferred the population parameters, which was implemented in a framework of a Bayesian hierarchical model (BHM). A BHM is statistical model that works on several levels, where lower levels inherit their probability distributions from higher levels. In this case there are two levels: the top-level population parameters $\popp$ that describe the overall population of stars, and the bottom-layer of stellar parameters $\objp_i$ that describe the properties of individual stellar objects. The stellar parameters are connected to the data $\data_i$.

From Bayes' theorem, the posterior on the population parameters, with stellar parameters marginalised, is proportional to
\begin{equation}\label{eq:Bayes}
	\text{Pr}(\popp \, | \, \data_{i=\{1,...,N\}},S) \propto\text{Pr}(\popp)
    \prod_{i=1}^N \int
    \frac{S(\data_i) \text{Pr}(\data_i \, | \, \objp_{i})\text{Pr}(\objp_{i} \, | \, \popp)}{\bar{N}(\popp,S)} \de\objp_{i},
\end{equation}
where $\text{Pr}(\popp)$ is the prior over the population parameters, $S(\data_i)$ is the selection function, $\text{Pr}(\data_i \, | \, \objp_{i})$ is the likelihood of the data given the stellar parameters, $\text{Pr}(\objp_{i} \, | \, \popp)$ is the probability of the stellar parameters given the population model with fixed population parameters, and $\bar{N}(\popp,S)$ is the normalisation to the nominator of the integrand. These terms are discussed in detail below.

The posterior over the population parameters is written with a proportionality sign. The quantity of interest is the ratio of posterior values between points in population parameter space. Therefore any constant factor that does not depend on the population parameters can be dropped.

\subsection{Prior}
\label{sec:prior}

The population parameter prior that we used in this work is uniform in the volume of parameter space that fulfil the following conditions:
\begin{equation}
\begin{split}
& 0~\Msunppcc<\rho_{\{1,2,3,4\}}<0.2~\Msunppcc, \\
& 0 < c_2 < 1, \\
& 0 < c_3 < 1, \\
& c_2 + c_3 < 1, \\
& 0~\kmsec<\sigma_1<\sigma_2<\sigma_3<200~\kmsec, \\
& -50~\pc < Z_\odot < 50~\pc, \\
& -30~\kmsec< W_\odot < 30~\kmsec,
\end{split}
\end{equation}
and is otherwise zero valued. The velocity dispersions $\sigma_j$ are required to be in ascending order for reasons of multiplicity, which would otherwise cause sampling difficulties.

\subsection{Integration over stellar parameters}
\label{sec:objp_integration}

When evaluating the posterior, as expressed in Eq.~\ref{eq:Bayes}, the most demanding part computationally is the integral over the stellar parameters,
\begin{equation}
	\int S(\data_i)\, \text{Pr}(\data_i \, | \, \objp_{i})\, \text{Pr}(\objp_{i} \, | \, \popp)\, \de\objp_{i},
\end{equation}
which is done $N$ times, where $N$ is the number of stellar objects in the sample. For a given stellar object, the data is fixed, meaning that the selection factor $S(\data_i)$ is a constant that can be dropped. Because uncertainties on $l$, $b$, and $m_G$ are neglected, the integral is reduced from seven to four dimensions, where these degrees of freedom are parametrised by distance and velocity in three dimensions.

The integral can in part be computed analytically, as the velocity distribution is a sum of multivariate Gaussians and uncertainties in velocity are described by a Gaussian covariance matrix. For this reason, it is only necessary to numerically compute the one-dimensional integration over distance.

Dropping the index $i$ on the right-hand side for shorthand, the integral can be written
\begin{equation}\label{eq:objp_analytical_split}
\begin{split}
	\int & \text{Pr}(\data_i \, | \, \objp_{i})\, \text{Pr}(\objp_{i} \, | \, \popp)\, \de\objp_{i} \propto \\
    \int & \Bigg[
    \int \text{Pr}(\data \, |\, \boldsymbol{v},\varpi)\, \text{Pr}(\boldsymbol{v}\, |\, \popp,\varpi) \, n(z)
    \, \de^3\boldsymbol{v}
    \Bigg]\\
    & \times\, \mathcal{G}(\varpi-\hat{\varpi},\hat{\sigma}_\varpi)\, F(M_G)\, d^2\,  \de d
\end{split}
\end{equation}
where the integral over velocity, within the large brackets, has an analytic form. It is implicit that the true parallax value $\varpi$ is given by the inverse of the distance $d$, and that the absolute magnitude $M_G$ is set by the distance $d$, apparent magnitude $m_G$, and dust extinction $\Xi(\boldsymbol{x})$. In principle, also the function $n(z)$ can be moved outside the inner integral, but not doing so makes the end result somewhat more elegant.

We express the likelihood of the data $\text{Pr}(\data \, |\, \boldsymbol{v},\varpi)$ in terms of velocities $v_l$, $v_b$, and $v_{RV}$, which are the velocities in longitudinal, latitudinal and radial directions. The integral over velocities is performed under a fixed true parallax value $\varpi$ (or fixed distance, equivalently). Quantities which are conditional on a fixed parallax are marked with a tilde. The conditional likelihood mean value for velocity in the latitudinal direction is equal to
\begin{equation}\label{eq:tilde_v}
	\tilde{v}_l = \frac{k_\mu}{\varpi} \,\left( \hat{\mu}_l+\hat{\rho}_{\varpi\mu_l}\hat{\sigma}_{\mu_l}\frac{\varpi-\hat{\varpi}}{\hat{\sigma}_\varpi}\right),
\end{equation}
where $k_\mu = 4.74057~\text{yr}\times\kmsec$ is a unit conversion constant. The equivalent expression holds for $\tilde{v}_b$. We define a vector of conditional mean values for the velocity likelihood, which is
\begin{equation}
	\tilde{\boldsymbol{v}} \equiv
    \begin{bmatrix}
    \tilde{v}_l \\
    \tilde{v}_b \\
    \hat{v}_{RV}
    \end{bmatrix}.
\end{equation}
The conditional error covariance matrix for velocities in the latitudinal and longitudinal directions is
\begin{equation}
\begin{split}
	& \tilde{\boldsymbol{\Sigma}}_\mu = \\
    & \begin{bmatrix}
	\dfrac{k_\mu}{\varpi}\hat{\sigma}_{\mu_l}^2(1-\hat{\rho}_{\varpi\mu_l}^2)
	& \dfrac{k_\mu}{\varpi}\hat{\sigma}_{\mu_l}\hat{\sigma}_{\mu_b}(\hat{\rho}_{\mu_l\mu_b}-\hat{\rho}_{\varpi\mu_l}\hat{\rho}_{\varpi\mu_b}) \\
	\dfrac{k_\mu}{\varpi}\hat{\sigma}_{\mu_l}\hat{\sigma}_{\mu_b}(\hat{\rho}_{\mu_l\mu_b}-\hat{\rho}_{\varpi\mu_l}\hat{\rho}_{\varpi\mu_b})
	& \dfrac{k_\mu}{\varpi}\hat{\sigma}_{\mu_b}^2(1-\hat{\rho}_{\varpi\mu_b}^2)
   \end{bmatrix}.
\end{split}
\end{equation}
The full conditional error covariance matrix for velocity is
\begin{equation}
	\tilde{\boldsymbol{\Sigma}} = 
    \begin{bmatrix}
		\tilde{\boldsymbol{\Sigma}}_\mu & \emptyset \\
        \emptyset & \hat{\sigma}_{RV}^2
   	\end{bmatrix},
\end{equation}
where $\emptyset$ are zero valued sub-matrices of shapes $1 \times 2$ and $2 \times 1$.

The velocity distribution $\text{Pr}(\boldsymbol{v}\, |\, \popp,\varpi)$ is described by a Gaussian mixture, according to
\begin{equation}
\text{Pr}(\boldsymbol{v}\, |\, \popp,\varpi)\, n(z) = 
	\sum_{k=1}^5 \, \sum_{j=1}^3\,
    a_k \, c_j \,
    \exp\left[-\frac{\Phi(Z)}{\sigma_j^2}\right] \,
    \mathcal{N}(\boldsymbol{v}-\boldsymbol{v}_{k},\boldsymbol{\Sigma}_{jk}),
\end{equation}
where
\begin{equation}
	\boldsymbol{v}_{k} = 
    \begin{bmatrix}
    	g_{u,k} \\
        g_{v,k} \\
        -W_\odot
    \end{bmatrix},
\end{equation}
and
\begin{equation}
	\boldsymbol{\Sigma}_{jk} = 
    \begin{bmatrix}
    	\boldsymbol{G}_{k} & \emptyset \\
        \emptyset & \sigma_j^2
    \end{bmatrix}.
\end{equation}

Because the likelihood of the data is also on a multivariate Gaussian form, the integral over velocity in Eq.~\eqref{eq:objp_analytical_split} becomes an analytic sum over the multivariate Gaussians of the velocity distribution. It is equal to
\begin{equation}\label{eq:including_RV}
\begin{split}
	\int & \text{Pr}(\data \, |\, \boldsymbol{v},\varpi)\, \text{Pr}(\boldsymbol{v}\, |\, \popp,\varpi) \, n(z)
    \, \de^3\boldsymbol{v} = \\
	& \sum_{k=1}^5 \, \sum_{j=1}^3\,
    a_k \, c_j \,
    \exp\left[-\frac{\Phi(Z)}{\sigma_j^2}\right]
    \mathcal{N}\big(\boldsymbol{R}\tilde{\boldsymbol{v}}-\boldsymbol{v}_{k},\boldsymbol{R}\tilde{\boldsymbol{\Sigma}}\boldsymbol{R}^{-1}+\boldsymbol{\Sigma}_{jk}\big),
\end{split}
\end{equation}
where
\begin{equation}
	\boldsymbol{R} =
    \begin{bmatrix}
    -\sin(l) &  -\cos(l)\sin(b) & \cos(l)\cos(b) \\
    \cos(l) & -\sin(l)\sin(b) & \sin(l)\cos(b)  \\
    0 & \cos(b) & \sin(b)
    \end{bmatrix}
\end{equation}
is the rotational matrix going from longitudinal-latitudinal-radial velocities to $u$-$v$-$w$ velocities.

In the case that an object is missing a radial velocity observation, the three-dimensional velocity distribution is projected to the plane perpendicular to the line-of-sight. The integral over velocity becomes
\begin{equation}\label{eq:not_including_RV}
\begin{split}
	\int & \text{Pr}(\data \, |\, \boldsymbol{v},\varpi)\, \text{Pr}(\boldsymbol{v}\, |\, \popp,\varpi) \, n(z)
    \, \de^3\boldsymbol{v} = \\
	& \sum_{k=1}^5 \, \sum_{j=1}^3\,
    a_k \, c_j \,
    \exp\left[-\frac{\Phi(Z)}{\sigma_j^2}\right] 
    \mathcal{N}\Big(
    \big\langle \tilde{\boldsymbol{v}}-\boldsymbol{R}^{-1}\boldsymbol{v}_{k} \big\rangle,
    \big\langle \tilde{\boldsymbol{\Sigma}}+\boldsymbol{R}^{-1}\boldsymbol{\Sigma}_{jk}\boldsymbol{R} \big \rangle
    \Big),
\end{split}
\end{equation}
where $\big\langle ... \big \rangle$ denotes this projection; for example, $\big\langle \tilde{\boldsymbol{v}} \big \rangle = [\tilde{v}_l,\tilde{v}_b]$ and $\big\langle \tilde{\boldsymbol{\Sigma}} \big \rangle = \tilde{\boldsymbol{\Sigma}}_\mu$.

It might seem superfluous to integrate over the velocity for stellar objects that are missing radial velocity measurements and are situated in the direction of the Galactic north or south pole, as they do not directly constrain the vertical velocity distribution. However, the proper motion is still informative about the distance of such a stellar object, since their velocity parallel to the plane must be consistent with $f_\parallel$. This is especially informative when the parallax uncertainty is large.

\subsection{Normalisation}
\label{sec:norm}

The normalisation $\bar{N}(\popp,S)$ is in the denominator of the integrand of Eq.~\eqref{eq:Bayes}. It normalises the nominator of the integrand, and depends on the population parameters $\popp$ and the selection function $S$. It is equal to
\begin{equation}\label{eq:norm}
	\bar{N}(\popp,S) =
    \int S(\data_x)\, \text{Pr}(\data_x \, | \, \objp_x)\, \text{Pr}(\objp_x \, | \, \popp)\, \de\objp_x\, \de\data_x.
\end{equation}
The index $x$ on $\objp_x$ and $\data_x$ signifies that this is not an integral over an object in the sample, but an integral over a hypothetical object's generated stellar parameters and data, as drawn from the population model.

The selection function, described in Sect.~\ref{sec:selection}, is only dependent on position and magnitude, where uncertainty on angles $l$ and $b$ are small enough to be neglected. Therefore, the integral in the normalisation factor reduces to
\begin{equation}\label{eq:norm2}
\begin{split}
	\bar{N}(\popp,S) =
    \int & S(\hat{\varpi},m_G,l,b)\,
    \text{Pr}(\hat{\varpi} \, | \, \boldsymbol{x},m_G) \\ \times\,
    & n(Z)\, F(M_G)\,
    \de^3\boldsymbol{x}\, \de M_G\, \de\hat{\varpi},
   \end{split}
\end{equation}
where the only degree of freedom that remains in terms of the data is the observed parallax. It is implicit that the apparent magnitude is given by the true position and true absolute magnitude, according to
\begin{equation}\label{eq:m_G}
	m_G=M_G+5\Bigg[\log_{10}\Bigg(\frac{\arcs}{\varpi}\Bigg)-1\Bigg]+\Xi(\boldsymbol{x}).
\end{equation}

The only component of Eq.~\eqref{eq:norm2} that remains to be explained is $\text{Pr}(\hat{\varpi} \, | \, \boldsymbol{x},m_G)$, which is the probability of observing a certain parallax, given a stellar object's true position and true absolute magnitude. It is given by an integral over parallax uncertainties,
\begin{equation}
	\text{Pr}(\hat{\varpi} \, | \, \boldsymbol{x},m_G) = \int
    \mathcal{G}(\varpi-\hat{\varpi},\hat{\sigma}_\varpi)\,
    h(\hat{\sigma}_\varpi\, |\, m_G,b)\,
    \de\hat{\sigma}_\varpi,
\end{equation}
where $h(\hat{\sigma}_\varpi\, |\, m_G,b)$ is the density of parallax uncertainties as described in Sect.~\ref{sec:parallax_errors}.

Despite the above simplifications, this integral is still high-dimensional and expensive to compute. However, it is possible to catalogue a vector of effective area $A_\text{eff}$, reducing this integral to only one dimension. The normalisation factor can be written on the form
\begin{equation}\label{eq:norm3}
	\bar{N}(\popp,S) = \int n(Z)\,A_\text{eff}(z)~\de z,
\end{equation}
where $n(Z)$ is the number density function of Eq.~\eqref{eq:nuofz}, and $Z=z+Z_\odot$. The effective area is given by
\begin{equation}\label{eq:effective_are}
\begin{split}
	A_\text{eff}(z) = \int S(\hat{\varpi},m_G,l,b)\,
    \mathcal{G}(\varpi-\hat{\varpi},\hat{\sigma}_\varpi)\,
    h(\hat{\sigma}_\varpi\, |\, m_G,b)\, \\
    \times \, F(M_G)\,
    \de x\, \de y\, \de M_G\, \de\hat{\sigma}_\varpi \, \de\hat{\varpi},
\end{split}
\end{equation}
where the true parallax $\varpi$ has an implicit dependence on the coordinates $\boldsymbol{x} = \{x,y,z\}$. The effective area does not depend on the population parameters, but only on fixed background distributions. The effective area is plotted in the control plots of Appendix~\ref{app:controlplots}.

\section{Results}\label{sec:res}

In order to explore the posterior distribution of the population parameters, as defined in Eq.~\eqref{eq:Bayes}, we sampled this function using a Metropolis-Hastings Markov chain Monte-Carlo (MCMC) algorithm \citep{BayesianDataAnalysis}. After a thorough burn-in phase, where the maximum mode was located and the step length is calibrated, the MCMC was run for 10,000 steps. The MCMC chain was then thinned by a factor of 10. The code that was used to generate the results presented in this article is open source and available online.\footnote{\url{https://github.com/AxelWidmark/densityDR2}}

\subsection{Mock data}
In order to test the method and its constraining power, the statistical method was applied to four mock data samples. The construction of mock data samples and the statistical results are explained in detail in Appendix~\ref{app:mockdata}. For all mock data samples, the matter distribution of the generated data was recovered in the 90 per cent highest posterior density region, with statistical uncertainties similar to those of the real data samples.

\subsection{Gaia data}
For the eight \emph{Gaia} data samples, S1--S8, the population parameter inference was performed for three separate cases. In the first case, we used the velocity information of all stellar objects. In the second case, we used the velocity information only of stellar objects with a good radial velocity measurement ($\hat{\sigma}_{RV}<3~\kmsec$, amounting to 75--85 per cent of the respective sample's total number of stellar objects), with no dependence of the horizontal velocity distribution $f_\parallel$. For the third case, we used the velocity information of stellar objects with $|b|\leq 5$ degrees, and therefore close to the mid-plane. For stellar objects where the velocity information was neglected, only their position was integrated over.

Comparing the results of these three cases functions as a test of possible systematic errors to the model, which could manifest as an inconsistency between them. The second case does not depend on the horizontal velocity distribution, providing a test of the approximation that $f_\parallel$ and $f_\perp$ are separable. The latter case is similar to the method used by \cite{WidmarkMonari,Schutz:2017tfp,Buch:2018qdr}, and is therefore useful for comparing results with these previous studies.

\begin{figure}
	\centering
	\includegraphics[width=\columnwidth]{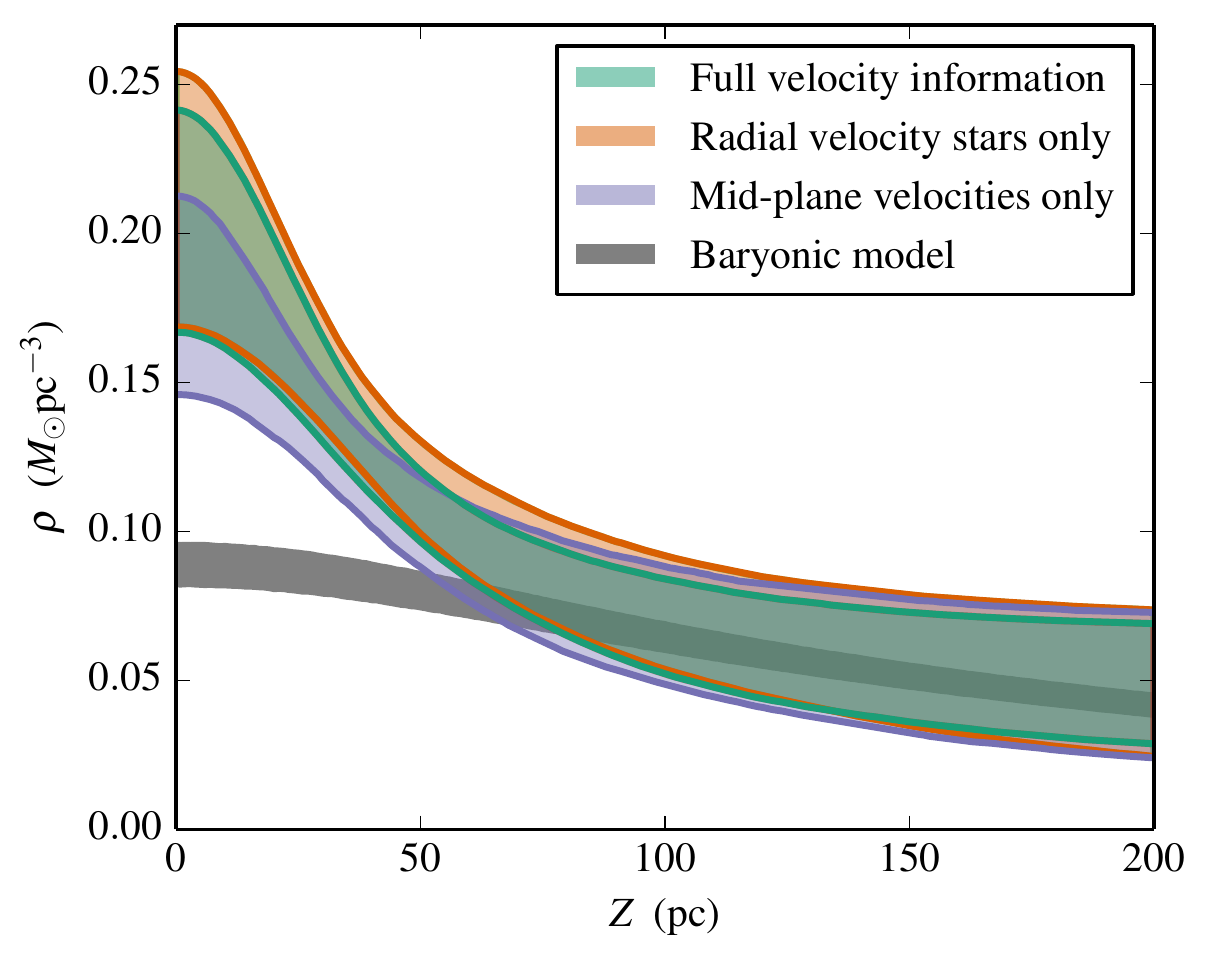}
    \caption{Inferred total matter density of data samples S1--S8, as a function of height with respect to the Galactic plane. The matter density is plotted for three different cases: when using the velocity information of all stellar objects, when using the velocity information of stellar objects with good radial velocity measurements only, and when using the velocity information of stellar objects in the mid-plane only. Also shown is the 1-$\sigma$ band of the baryonic model (see Sect.~\ref{sec:baryons}).}
    \label{fig:density_S1-S8}
\end{figure}

The inferred density of samples S1--S8 are shown in Fig. \ref{fig:density_S1-S8}, for the three cases described above. The respective bands correspond to the 16th and 84th percentiles of the inferred density at different heights $Z$. In this figure, the eight samples' posterior densities are summed together; hence, the width of the bands include statistical variance within samples as well as variance between samples. Also shown is a baryonic model, as described in Sect.~\ref{sec:baryons}. The density at heights $|Z| \gtrsim 60~\pc$ is consistent with the baryonic model and a dark matter halo density of $0.01~\Msunppcc$ (although not strong enough to constrain it). Closer to the mid-plane, the inferred matter density is significantly higher than that of the baryonic model, with a surplus that is roughly in the range $0.05$--$0.15~\Msunppcc$. The case with full velocity information agrees well with the case where we consider the velocity information only of stellar objects with good radial velocity measurements, although the inferred densities are marginally higher. The case with only mid-plane velocities differs somewhat, preferring a lower mid-plane matter density for most samples. The inferred matter density is also shown in more detailed figures in Appendix~\ref{app:data}.

In Fig.~\ref{fig:surplus}, we show the surplus of the inferred matter density distribution within 60 pc from the Galactic disc, for the three cases described above. The surplus is with respect to a background model, consisting of a homogeneous dark matter halo with a density of $0.01~\Msunppcc$ and a distribution of baryons, following the model described Sect.~\ref{sec:baryons} and using the mean values of Table~\ref{tab:baryonic_components}. The uncertainty of the matter density of the baryonic model is $0.0071~\Msunppcc$ in the mid-plane. Assuming that the dark matter halo uncertainty is sub-dominant, this amounts to an uncertainty of about $0.85~\Msunppcsquare$ to the surface density within $|Z|<60~\pc$.

For the first case where the velocity information of all stellar objects are accounted for (upper panel), samples S1--S5 agree very well, centred on about $5~\Msunppcsquare$. Samples S6--S8, which are the least luminous, are outliers in the sense that they all infer higher mid-plane matter densities, possibly indicating some kind of systematic bias. The results of the second case, which has no dependence on the horizontal velocity distribution, shows a similar result. Comparing with the first case, the posterior medians are marginally higher and the posterior widths are inflated by up to 20 per cent. This indicates that the approximation of separability of $f_\parallel$ and $f_\perp$ does not give rise to any significant bias. For the third case where only mid-plane velocities are accounted for (lower panel), samples S6--S8 are not strong outliers, although they infer slightly higher values than most other samples. Compared to other two cases, the posterior widths are larger and the inferred surplus density is slightly lower for most samples.

In Appendix~\ref{app:controlplots} and Figs.~\ref{fig:control_S1}--\ref{fig:control_S8}, we show comparisons between model and binned data, and the residual between the two, for data samples S1--S8. It is clear that there is substructure in phase-space that our model does not account for (also seen in Fig.~\ref{fig:uvGMM}). There is no signature of asymmetry to the stellar number density across the Galactic plane (although statistics are low for $|z|\gtrsim 150$ pc), but there are signs of spatial substructure in all samples (left panels), where the residuals are as high as roughly three standard deviations at most. More severe are the substructures seen in the velocity distributions (right panels), where the highest residual is roughly four and a half standard deviations.

\begin{figure}
	\centering
	\includegraphics[width=\columnwidth]{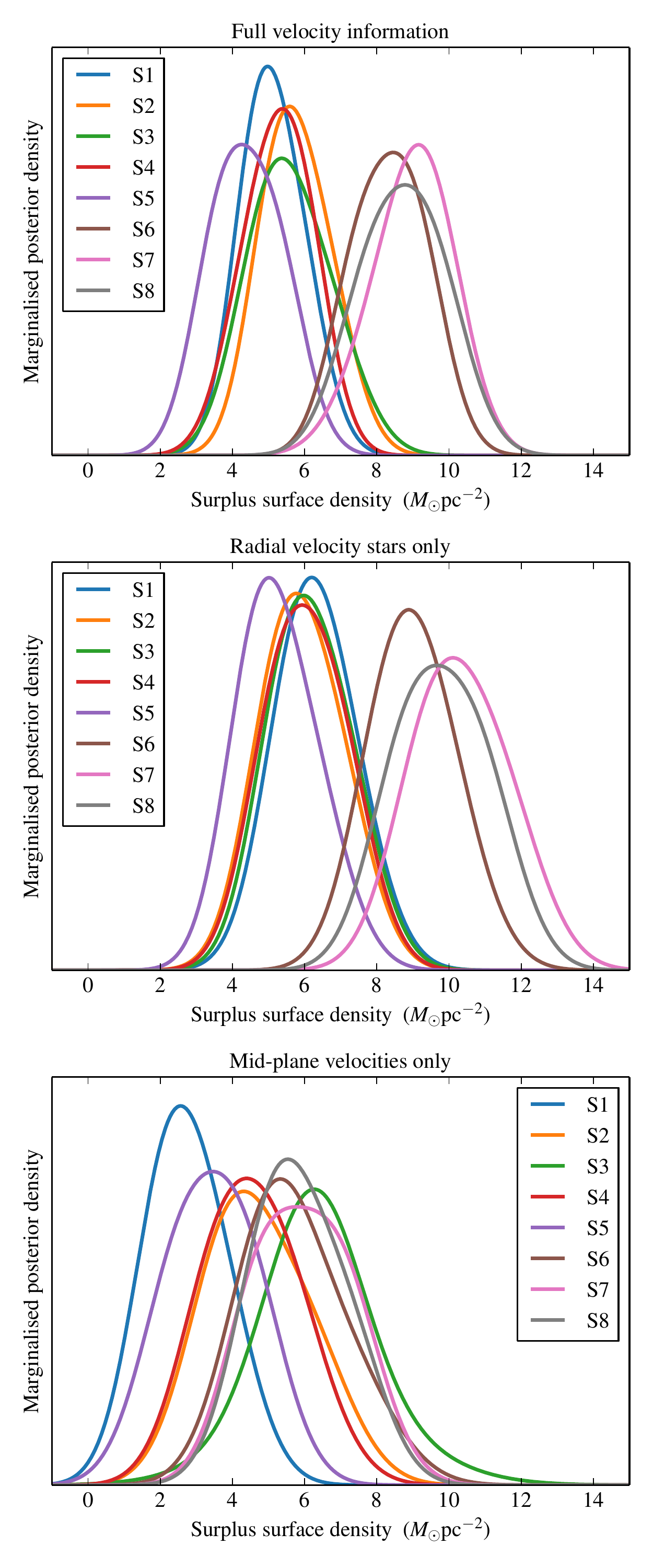}
    \caption{Surplus surface density of matter within $|Z|<60~\pc$, when comparing the inferred dynamical matter density with a model consisting of baryons and a dark matter halo. The upper panel shows the posterior for the case when the velocity information of all stellar objects are considered, the middle panel shows the case when only the velocity information of stellar objects with good radial velocity measurements are considered, while the lower panel is for the case when only mid-plane velocities are considered.}
    \label{fig:surplus}
\end{figure}

{\renewcommand{\arraystretch}{1.6}
\begin{table}
\caption{Inferred height $Z_\odot$ and vertical velocity $W_\odot$ of the Sun, with respect to the Galactic plane, for data samples S1--S5.}
\label{tab:pos_vel_sun}      
\centering          
\begin{tabular}{l l l} 
\hline
Sample & $Z_\odot$ (pc) & $W_\odot$ ($\kmsec$)\\
\hline
S1 & $3.86\pm 1.81$ & $7.30\pm 0.10$ \\
S2 & $7.06\pm 1.85$ & $7.32\pm 0.10$ \\
S3 & $4.90\pm 1.87$ & $7.45\pm 0.10$ \\
S4 & $4.55\pm 1.94$ & $7.42\pm 0.10$ \\
S5 & $3.46\pm 1.96$ & $7.17\pm 0.10$ \\
S6 & $4.87\pm 1.77$ & $7.14\pm 0.10$ \\
S7 & $5.38\pm 1.79$ & $7.26\pm 0.11$ \\
S8 & $2.64\pm 1.59$ & $6.86\pm 0.10$ \\
\hline
All samples & $4.59\pm 2.21$ & $7.24\pm 0.20$ \\
\hline
\end{tabular}
\end{table}}

The inferred height and vertical velocity of the Sun with respect to the Galactic plane ($Z_\odot$ and $W_\odot$) are presented in Table~\ref{tab:pos_vel_sun} (for the case when all velocity information is accounted for). The marginalised posteriors over these parameters agree well with a Gaussian distribution, which is why they are presented in terms of their mean values and standard deviation. For the inferred height of the Sun above the Galactic plane, the most severe tension is almost two standard deviations between samples S2 and S8. For the inferred vertical velocity of the Sun, the different samples are in reasonable agreement with the exception of S8; between S3 and S8 there is a tension of four standard deviations. This is suggestive of a systematic error in how the velocity distribution is modelled. These discrepancies between samples are not, in and of themselves, indicative of systematic errors in other parameters; the uncertainties are very small, and there are no strong correlations between $Z_\odot$, $W_\odot$ and other population parameters. For the values where all samples are considered in unison, the posteriors of the respective samples are added together (not multiplied); this is very conservative, in the sense that the uncertainties are inflated, also accounting for systematic discrepancies between samples.

While uncertainties of the dust map were not accounted for in this work, dust extinction probably has a negligible effect, as argued in Sect.~\ref{sec:dust}. We performed inference while completely ignoring dust effects, and found similar results as presented above.

\section{Discussion}\label{sec:conclusion}

In all eight stellar samples analysed in this work, we infer a surplus of matter confined to the Galactic plane ($|Z|\lesssim 60~\pc$), to high statistical significance. Comparing this with the expected density distribution of a baryonic model and a dark matter halo density of approximately $0.01~\Msunppcc$, this surplus corresponds to an excess surface density of approximately 5--9 $\Msunppcsquare$.

These results are in mild tension with \cite{Schutz:2017tfp} and \cite{Buch:2018qdr}, who report 95 \% constraints of around $7~\Msunppcsquare$ to a surplus matter density with a scale height of 30 pc. There are a number of differences between these two studies and our analysis. They only use the velocity information of stars close to the Galactic plane, and have fewer and smaller stellar samples. In the latter study, bases on \emph{Gaia} DR2, the limit comes from a sample of 4544 A type stars, out of which 310 constrain the velocity distribution.

There are some discrepancies between the results of the respective samples. While samples S1--S5 agree very well, samples S6--S8 are outliers with higher mid-plane matter densities. This is indicative of a smaller systematic error (not large enough to invalidate the inferred mid-plane surplus), likely correlated with stellar luminosity. As argued in Sect. \ref{sec:data}, this is unlikely to be due to systematic errors in the data (astrometric systematics are small, and the completeness function would have to be incorrect by at least a few per cent in order to explain this discrepancy). Potentially, the discrepancy is due to substructures in phase-space, which could be correlated with stellar type, luminosity and age, in the sense that different stellar populations are affected differently by non-stationary effects.

The inferred surplus matter density in the Galactic mid-plane could be explained by an underestimated density of baryons. The scale height of the matter surplus corresponds well with that of molecular gas, which is the coldest baryon component. It would be necessary to add a mid-plane density of at least $0.05~\Msunppcc$ to the cold gas components in the baryonic model, thereby doubling the matter density of cold gas. In the baryonic model, the statistical uncertainty of the combined cold atomic and molecular mid-plane densities amount to $0.0064~\Msunppcc$. In studies to determine the local density of gas, there is some variation comparable to the statistical error of the baryonic model used in this work \citep[see for example table 1 in][]{0004-637X-829-2-126}. The local density of cold gas would have to be severely misunderstood in order to account for the inferred surplus. As mentioned in Sect.~\ref{sec:intro}, an alternative and more exotic explanation for the surplus matter density could be double-disc dark matter, formed from a dark matter sub-component with strong dissipative self-interaction \citep{Fan:2013tia,Fan:2013yva}.

The results for the vertical velocity of the Sun with respect to the Galactic plane is in accordance with most previous studies, for example \cite{2010MNRAS.403.1829S,WidmarkMonari}. The results for the height of the Sun with respect to the Galactic plane ($4.76\pm 2.27$ pc) agree well with \cite{Buch:2018qdr}, but is in tension with for example \cite{Juric:2005zr,2017MNRAS.468.3289Y,WidmarkMonari,BovyAssym}, who report higher values of $25\pm 5~\pc$, $13.4\pm 4.4~\pc$, $15.3\pm 2.2~\pc$, and $20.8\pm 0.3~\pc$, respectively. \cite{BovyAssym} demonstrate that the distribution of stars is not symmetric across the Galactic plane, and exhibit wave-like patterns in both number density and vertical velocity, especially at greater distances from the plane (>200 pc). For this reason, measurements of the Sun's position with respect to the mid-plane can differ depending on how the mid-plane is defined, and what distance cuts are made in the analysis. \cite{Juric:2005zr,2017MNRAS.468.3289Y,BovyAssym} all extend to kpc distances and fit symmetric stellar number density distributions, which could explain why they all infer higher values.

As discussed in the beginning of Sect.~\ref{sec:galacticmodel}, there are phase-space structures observed with \emph{Gaia} that our model does not account for. In a first step to relax the assumptions of a well-mixed, symmetric, separable and decoupled vertical velocity distribution, we plan to extend this work by instead fitting a Gaussian mixture model to the full three-dimensional velocity distribution. Effective sampling of such a model would constitute a significant computational challenge, due to its high dimensionality. It should be feasible by implementing such a model in a computational framework capable of auto-differentiation and using Hamiltonian Monte-Carlo sampling.

This work adds to the increasingly complex picture of Galactic dynamics that is emerging with \emph{Gaia} DR2, calling for more sophisticated methods and models. With future data releases, increasingly accurate and robust dynamical determinations of the distribution of matter in the Galactic disc is expected.

\begin{acknowledgements}
I would like to thank Hiranya Peiris, Daniel Mortlock, Boris Leistedt, Jens Jasche, Sofia Sivertsson, and Giacomo Monari for helpful discussions. I would also like to thank the anonymous referee for contributing to this article. This work has made use of data from the European Space Agency (ESA) mission \emph{Gaia} (\url{https://www.cosmos.esa.int/gaia}), processed by the \emph{Gaia} Data Processing and Analysis Consortium (DPAC,
\url{https://www.cosmos.esa.int/web/gaia/dpac/consortium}). Funding for the DPAC has been provided by national institutions, in particular the institutions participating in the \emph{Gaia} Multilateral Agreement.
\end{acknowledgements}




\bibliographystyle{aa} 
\bibliography{thisbib} 

\begin{appendix} 

\section{Gaussian mixture model of the horizontal velocity distribution}\label{app:uvGMM}

The distribution of velocities parallel to the plane, $f_\parallel(u,v)$, was fitted using a Gaussian mixture model consisting of five Gaussians. It was fitted to \emph{Gaia} DR2 with the following data cuts: an observed distance in range 100--200 pc; an observed absolute magnitude in range 3.0--6.3; an available radial velocity value; a parallax uncertainty smaller than 0.1 mas. These cuts give a total of 165,593 stellar objects.

The fit was performed using \textsc{scikit-learn}. The numerical values of the five Gaussians of the mixture are presented in Table~\ref{tab:GMMvalues}.

{\renewcommand{\arraystretch}{1.6}
\begin{table}
\caption{Gaussians mixture model component values of $f_\parallel(u,v)$: weights $a_k$, mean values $\boldsymbol{g}_k$, and covariance matrices $\boldsymbol{G}_k$, where $k=\{1,2,3,4,5\}$.}             
\label{tab:GMMvalues}      
\centering          
\begin{tabular}{l l l  } 
\hline
$a_k$ & $\boldsymbol{g}_k$ (km/s) & $\boldsymbol{G}_k$ (km$^2$/s$^2$) \\ 
\hline
   0.3069
   & $\begin{bmatrix}
   -4.507 \\
   -8.793
   \end{bmatrix}$ 
   & $\begin{bmatrix}
   619.22 & 109.32 \\
   109.32 & 157.74
   \end{bmatrix}$ \\
   0.2830
   & $\begin{bmatrix}
   -19.516 \\
   -15.941
   \end{bmatrix}$ 
   & $\begin{bmatrix}
   487.11 & 102.97 \\
   102.97 & 415.84
   \end{bmatrix}$ \\
   0.1954
   & $\begin{bmatrix}
   -37.959 \\
   -39.759
   \end{bmatrix}$ 
   & $\begin{bmatrix}
   1202.66 & -39.76 \\
   -39.76 & 743.94
   \end{bmatrix}$ \\
   0.1836
   & $\begin{bmatrix}
   26.732 \\
   -24.433
   \end{bmatrix}$ 
   & $\begin{bmatrix}
   891.79 & -2.29 \\
   -2.29 & 352.43
   \end{bmatrix}$ \\
   0.0310
   & $\begin{bmatrix}
   -11.800\\
   -75.144
   \end{bmatrix}$ 
   & $\begin{bmatrix}
   6203.66 & -303.22 \\
   -303.22 & 4708.95
   \end{bmatrix}$ \\
\hline                  
\end{tabular}
\end{table}}

\section{Posterior densities}\label{app:data}

\begin{figure*}
	\centering
	\includegraphics[width=1.\linewidth]{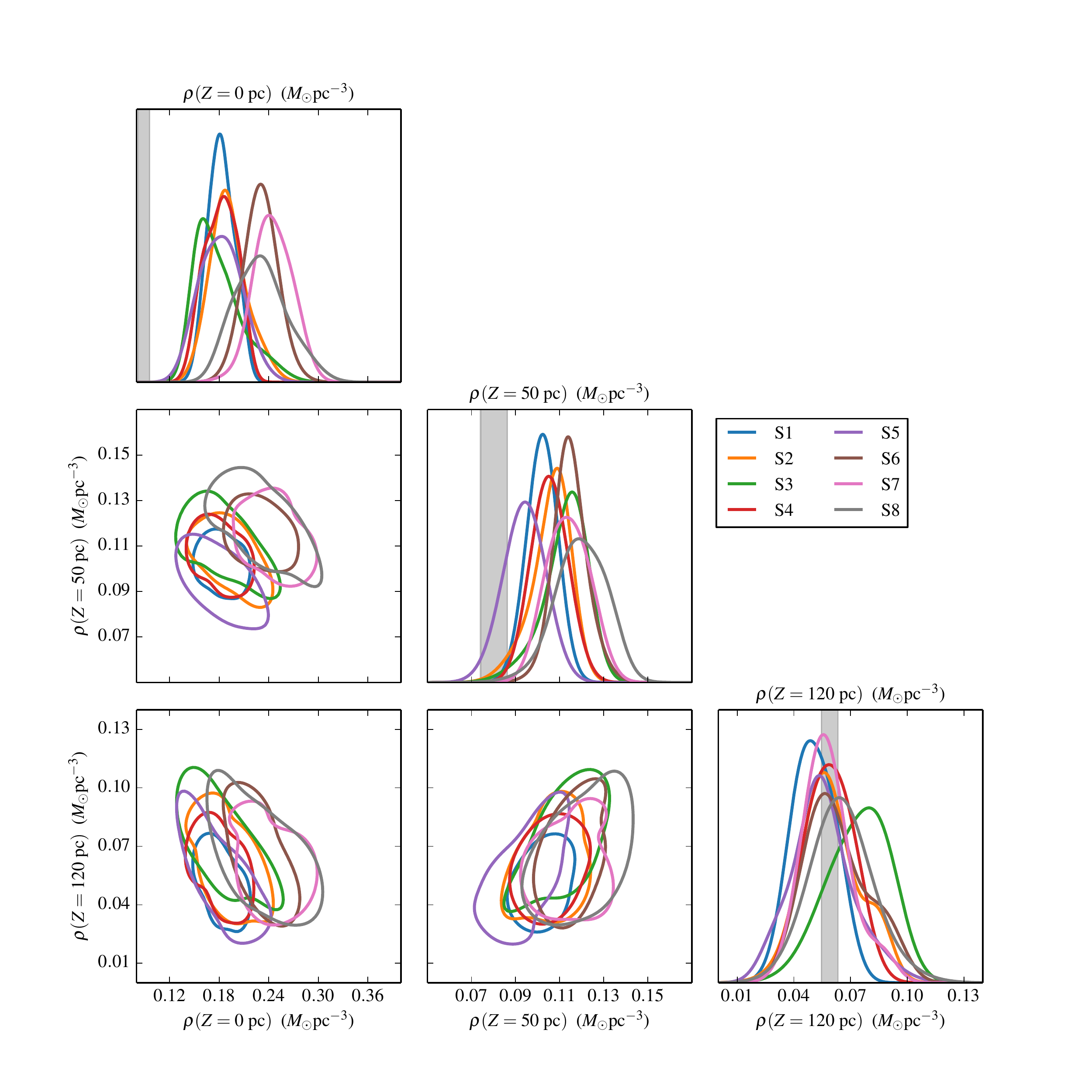}
    \caption{Inferred total dynamical matter density of data samples S1--S8 at heights 0, 50, and 120 pc above the Galactic plane, for the case where the velocity information of all stellar objects is accounted for. The panels on the diagonal show one-dimensional projections of the posterior probability density of the eight samples, as well as the 1-$\sigma$ band of the baryonic model in light grey. The other three panels show two-dimensional projections of the posteriors, represented as 90 per cent highest posterior density regions. Axis are shared between panels, apart from the vertical axis of the one-dimensional projections. The legend applies to all panels.}
    \label{fig:corner8}
\end{figure*}

\begin{figure*}
	\centering
	\includegraphics[width=1.\linewidth]{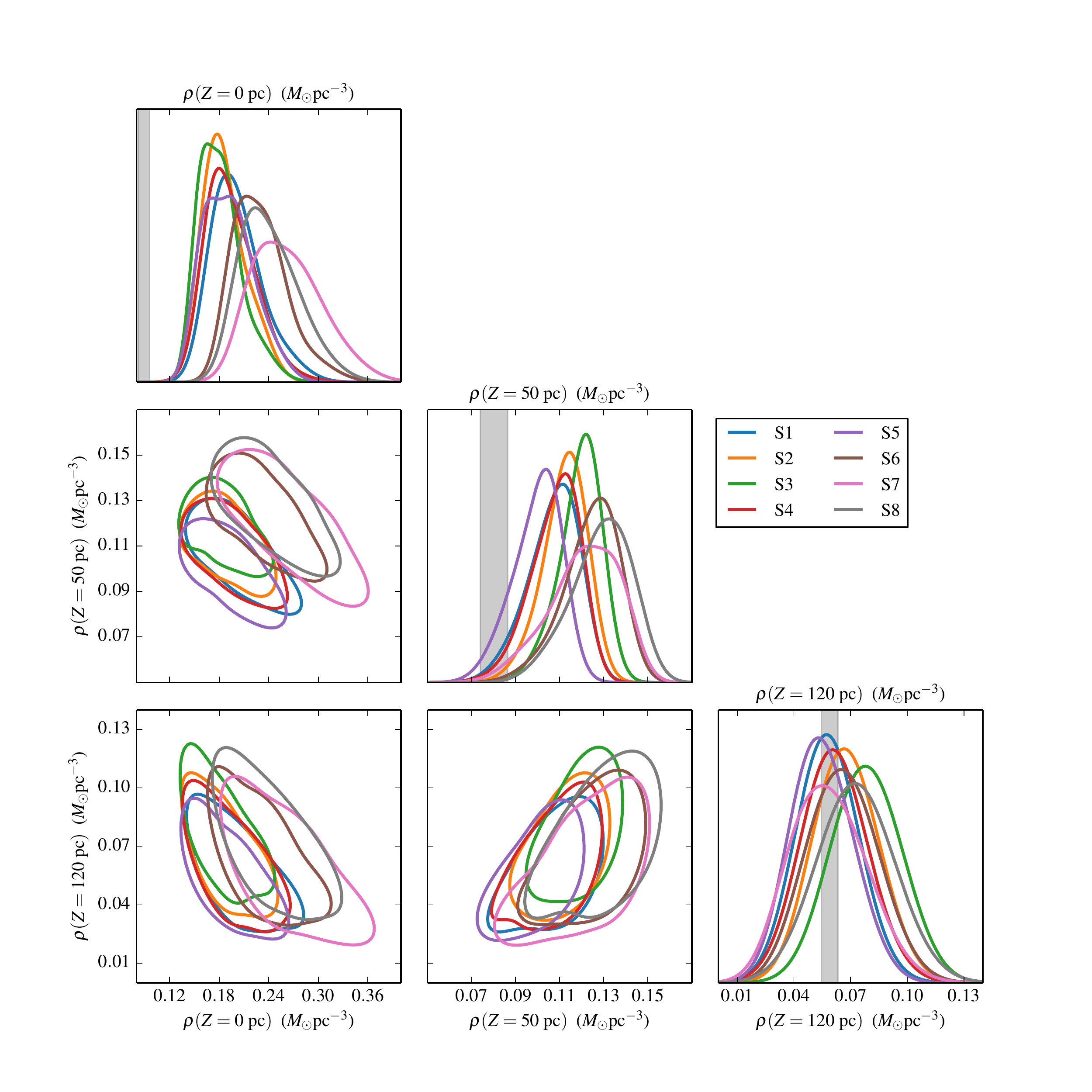}
    \caption{Inferred total dynamical matter density of data samples S1--S8 at heights 0, 50, and 120 pc above the Galactic plane, for the case where the velocity information of only the stellar objects with good radial velocity information is accounted for. The panels on the diagonal show one-dimensional projections of the posterior probability density of the eight samples, as well as the 1-$\sigma$ band of the baryonic model in light grey. The other three panels show two-dimensional projections of the posteriors, represented as 90 per cent highest posterior density regions. Axis are shared between panels, apart from the vertical axis of the one-dimensional projections. The legend applies to all panels.}
    \label{fig:corner8_clean}
\end{figure*}

\begin{figure*}
	\centering
	\includegraphics[width=1.\linewidth]{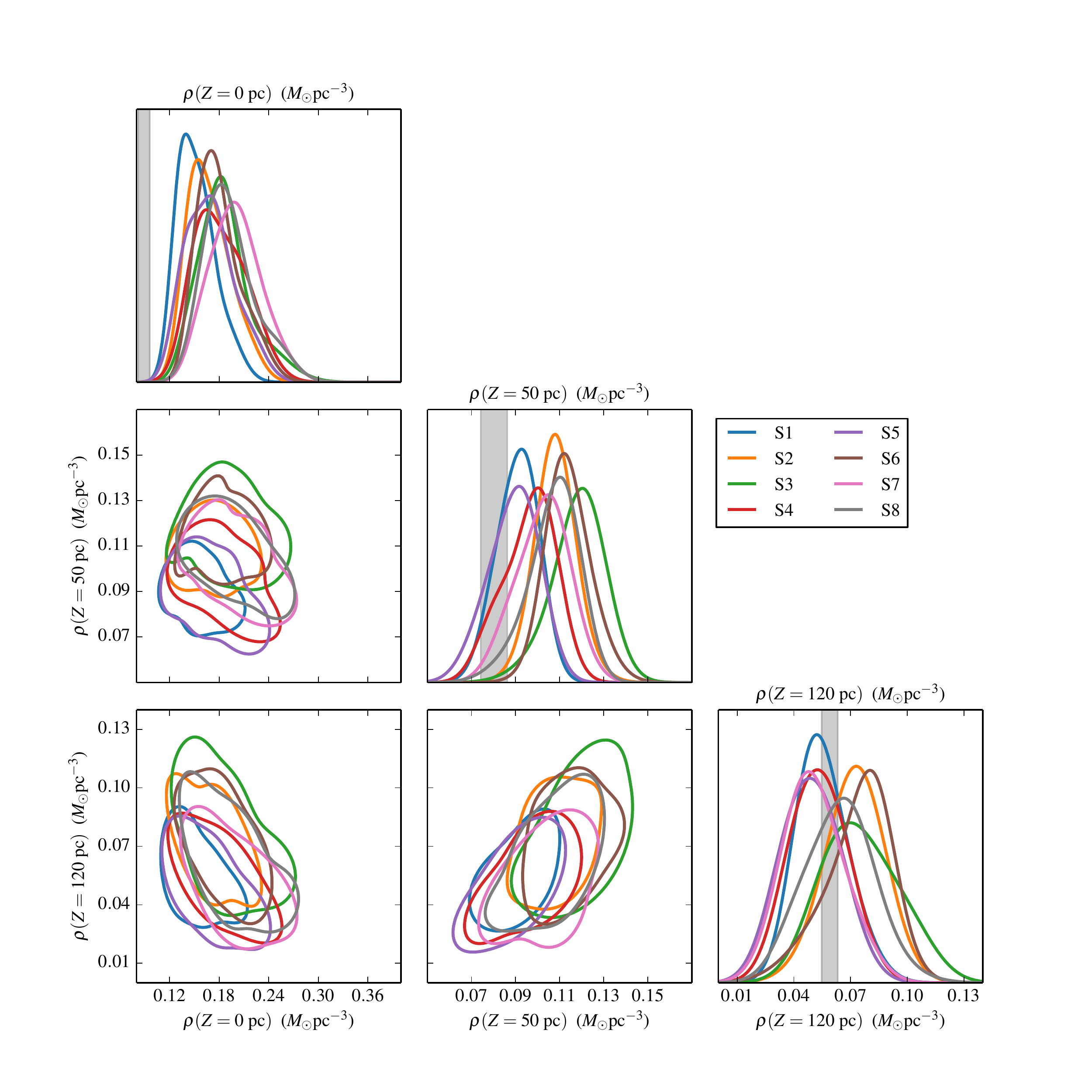}
    \caption{Inferred total dynamical matter density of data samples S1--S8 at heights 0, 50, and 120 pc above the Galactic plane, for the case where the velocity information of only the mid-plane stellar objects is accounted for. The panels on the diagonal show one-dimensional projections of the posterior probability density of the eight samples, as well as the 1-$\sigma$ band of the baryonic model in light grey. The other three panels show two-dimensional projections of the posteriors, represented as 90 per cent highest posterior density regions. Axis are shared between panels, apart from the vertical axis of the one-dimensional projections. The legend applies to all panels.}
    \label{fig:corner8_in_plane}
\end{figure*}

The inferred matter density of samples S1--S8 is shown in Fig.~\ref{fig:corner8}, for the case when all velocity information is used, in Fig.~\ref{fig:corner8_clean}, where only where only the velocity information of stellar objects with radial velocity measurements are used (no dependence on the horizontal velocity distribution), and in Fig.~\ref{fig:corner8_in_plane}, where only the velocity information of the mid-plane stellar objects are used. These posterior probability distributions are not presented in terms of the population parameters $\rho_{\{1,2,3,4\}}$, but in terms of the inferred matter density at $Z=\{0,50,120\}$ pc (which is a linear combination of $\rho_{\{1,2,3,4\}}$). In order to facilitate an easy comparison, all the axis ranges are the same for these three plots, for each respective panel.

In all three figures, the density close to the mid-plane, at $Z=0$ pc, is somewhat anti-correlated with the densities at $Z=50$ pc and $Z=120$ pc. In Figs.~\ref{fig:corner8} and \ref{fig:corner8_clean}, there is tension between samples for the projection over posterior projection over $\rho(Z=0~\pc)$ and $\rho(Z=50~\pc)$; the 90 per cent highest posterior density region of sample S5 is barely or not at all overlapping with those of S6, S7, and S8. The statistical uncertainty is larger for Fig.~\ref{fig:corner8_in_plane}, and the posterior densities of the different samples overlap without apparent outlier groups.


\section{Mock data}\label{app:mockdata}
Four mock data samples named M1--M4 were generated using the following population parameter values:
\begin{equation}
    \begin{split}
    \rho_{\{1,2,3,4\}} & = \{0.05,0.05,0.03,0.03\}~\Msunppcc, \\
    c_2 & = 0.5, \\
    c_3 & = 0.05, \\
    \sigma_j & = \{9,20,50\}~\kmsec, \\
    Z_\odot & = 0~\pc, \\
    W_\odot & = 7.2~\kmsec.
    \end{split}
\end{equation}
The background distributions used to generate them were the same as those used for the inference on actual data. The sample cuts were the same as the real data samples S2--S5, listed in Table~\ref{tab:samples}.

The mock samples were constructed by rejection sampling, by generating objects from the true underlying model, assigning uncertainties and errors to those stellar objects, and including them in the sample if they meet the selection criteria. We took care to generate objects also outside the region defined by the selection cuts, to allow for stellar objects to scatter into the sample via observational errors.

The uncertainties on the parallax measurements were drawn from the background distribution $h(\hat{\sigma}_\varpi \, | \, m_G,b)$. The uncertainties on proper motions were drawn from corresponding distributions, given by data samples S2--S5. In the \emph{Gaia} data, the astrometric uncertainties are strongly correlated. We assumed maximum correlation for the mock data, such that the parallax uncertainty and the proper motion uncertainties of a stellar object were drawn from the same percentile in their respective distributions. We took the correlations between astrometric observables to be drawn at random these Gaussian distributions: $\mathcal{G}(0,0.22)$ for $\hat{\rho}_{\mu_l\varpi}$ and $\hat{\rho}_{\mu_b\varpi}$, and $\mathcal{G}(\hat{\rho}_{\mu_l\varpi} \hat{\rho}_{\mu_b\varpi},0.25)$ for $\hat{\rho}_{\mu_l\mu_b}$, which mimics the behaviour of correlations in the actual data.

The results of the inference on mock data is presented in Fig.~\ref{fig:corner4_mock}, in plots similar to those of Appendix~\ref{app:data}. Similarly to the case of the \emph{Gaia} data samples S1--S8, the density close to the mid-plane is anti-correlated with the density far from the Galactic plane.

\begin{figure*}
	\centering
	\includegraphics[width=1.\linewidth]{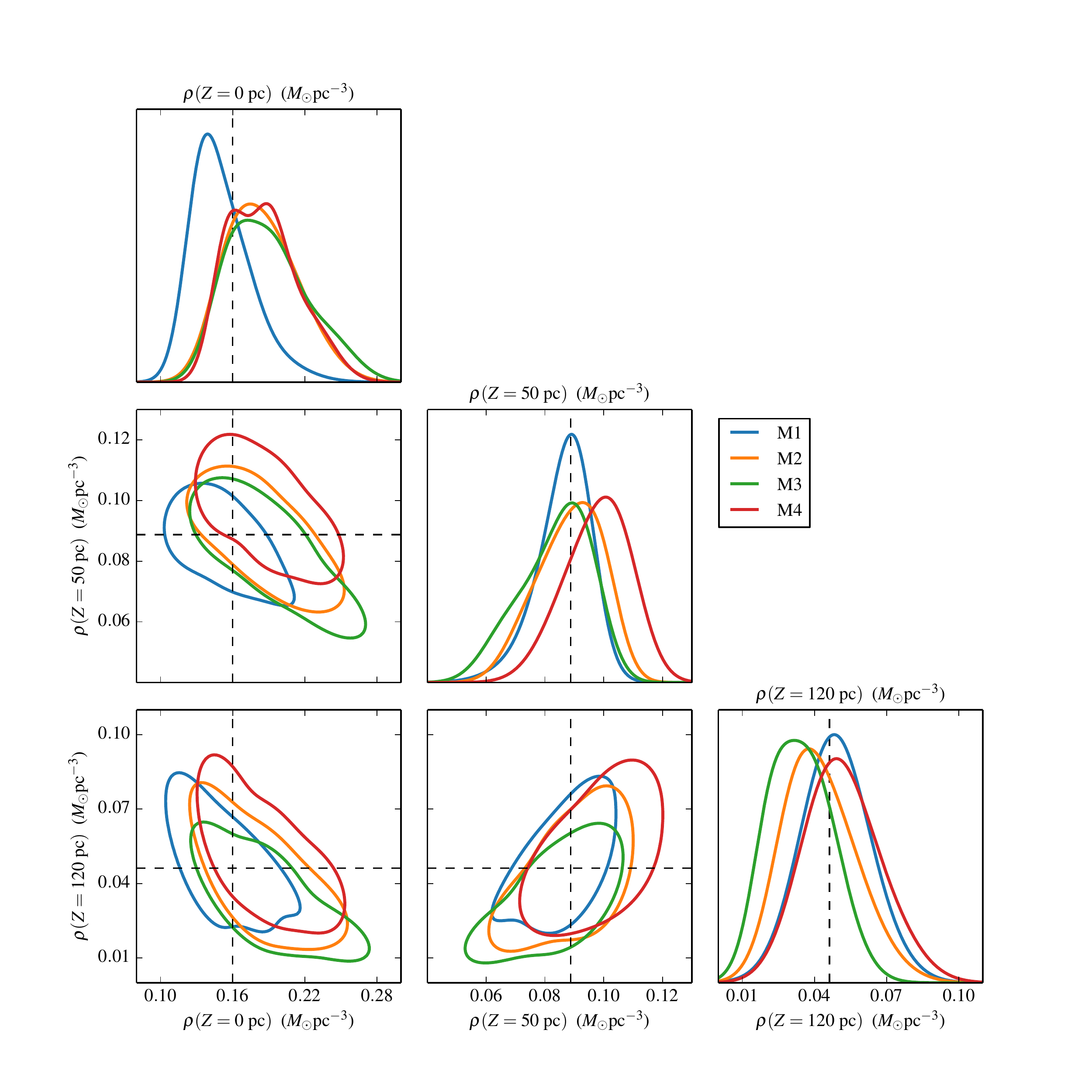}
    \caption{Inferred total dynamical matter density of mock data samples M1-M4, at heights 0, 50, and 120 pc above the Galactic plane. The panels on the diagonal show one-dimensional projections of the posterior probability density of the eight samples. The other three panels show two-dimensional projections of the posteriors, represented as 90 per cent highest posterior density regions. The dashed lines indicate the matter densities of the model used to generate the data. Axis are shared between panels, apart from the vertical axis of the one-dimensional projections. The legend applies to all panels.}
    \label{fig:corner4_mock}
\end{figure*}

\section{Control plots}\label{app:controlplots}

\begin{figure*}
	\centering
	\includegraphics[width=1.\linewidth]{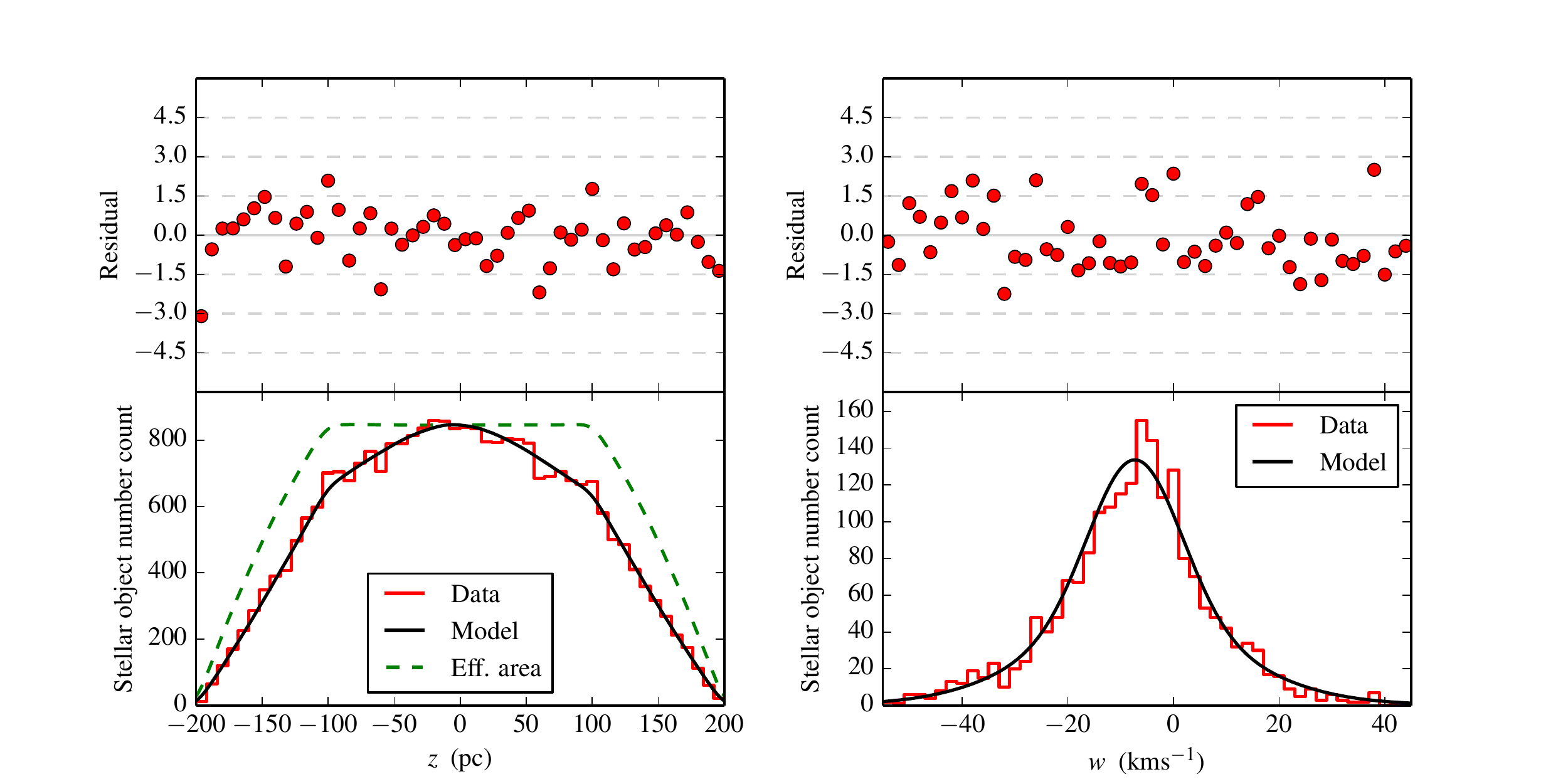}
    \caption{Control plot for sample S1, comparing the inferred model with the data. The left panel shows the stellar number count as a function of height with respect to the Sun. The right panel shows the vertical velocity distribution for stars close to the Galactic plane.}
    \label{fig:control_S1}
\end{figure*}

\begin{figure*}
	\centering
	\includegraphics[width=1.\linewidth]{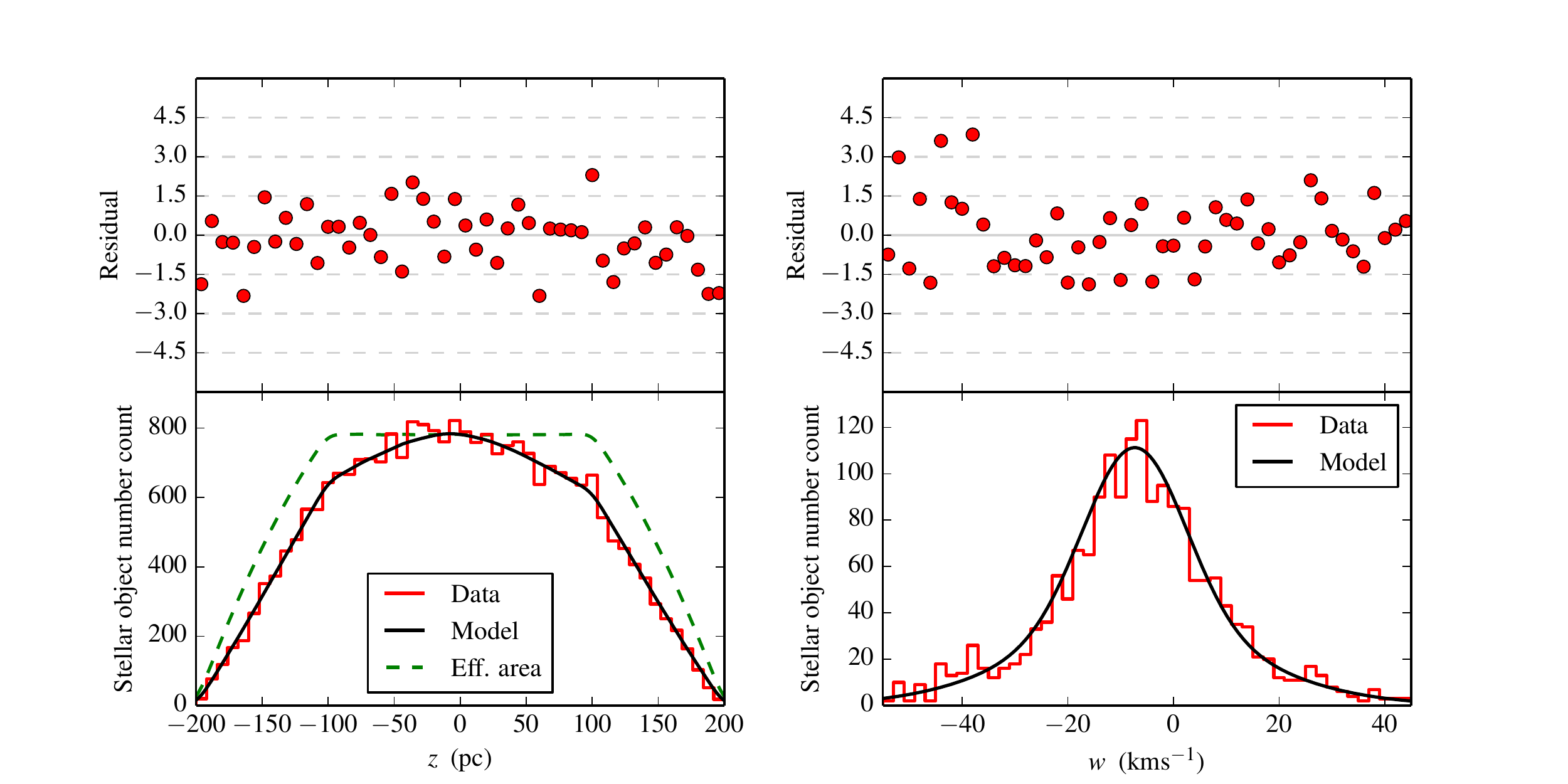}
    \caption{Control plot for sample S2, comparing the inferred model with the data. The left panel shows the stellar number count as a function of height with respect to the Sun. The right panel shows the vertical velocity distribution for stars close to the Galactic plane.}
    \label{fig:control_S2}
\end{figure*}

\begin{figure*}
	\centering
	\includegraphics[width=1.\linewidth]{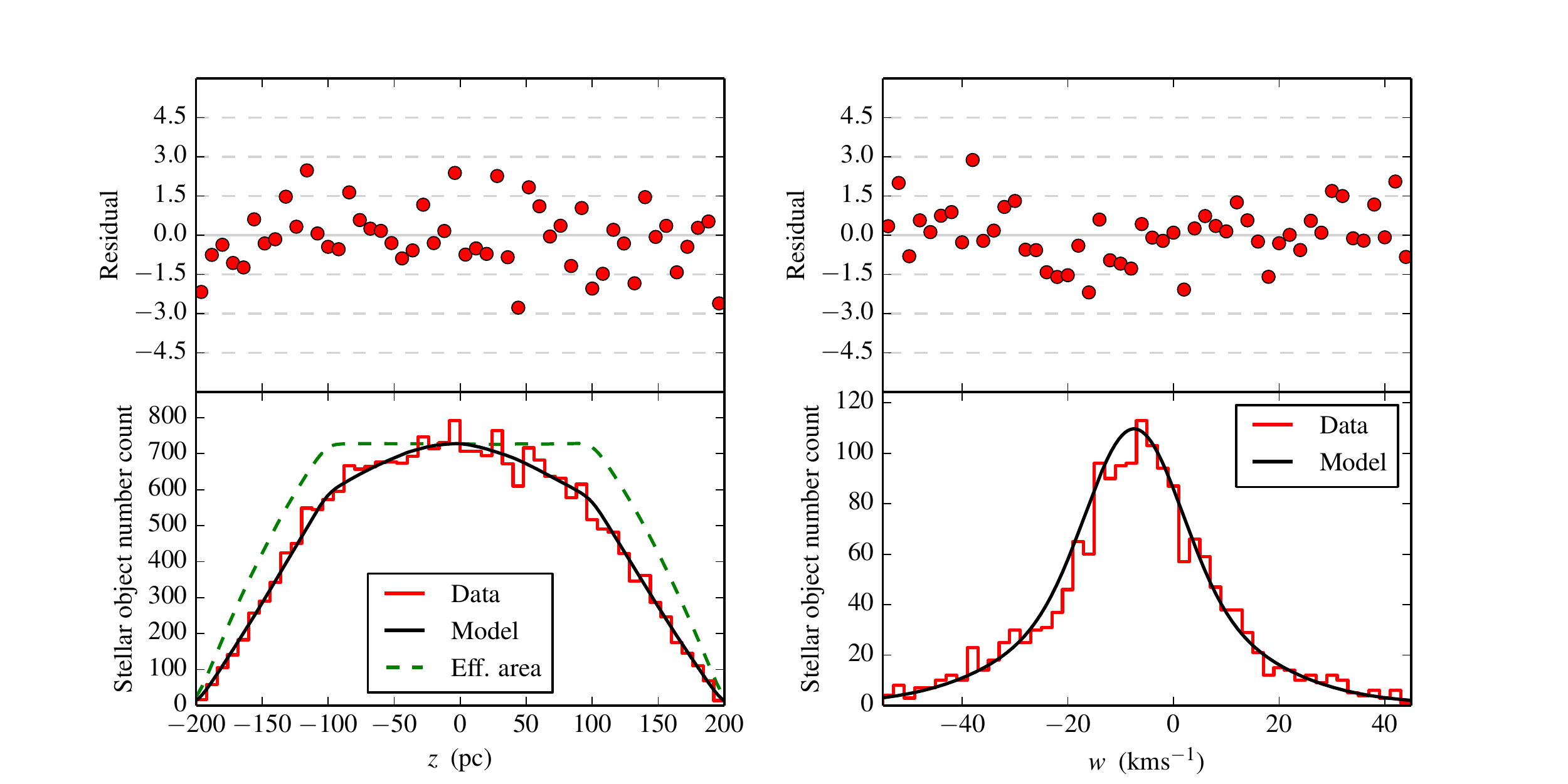}
    \caption{Control plot for sample S3, comparing the inferred model with the data. The left panel shows the stellar number count as a function of height with respect to the Sun. The right panel shows the vertical velocity distribution for stars close to the Galactic plane.}
    \label{fig:control_S3}
\end{figure*}

\begin{figure*}
	\centering
	\includegraphics[width=1.\linewidth]{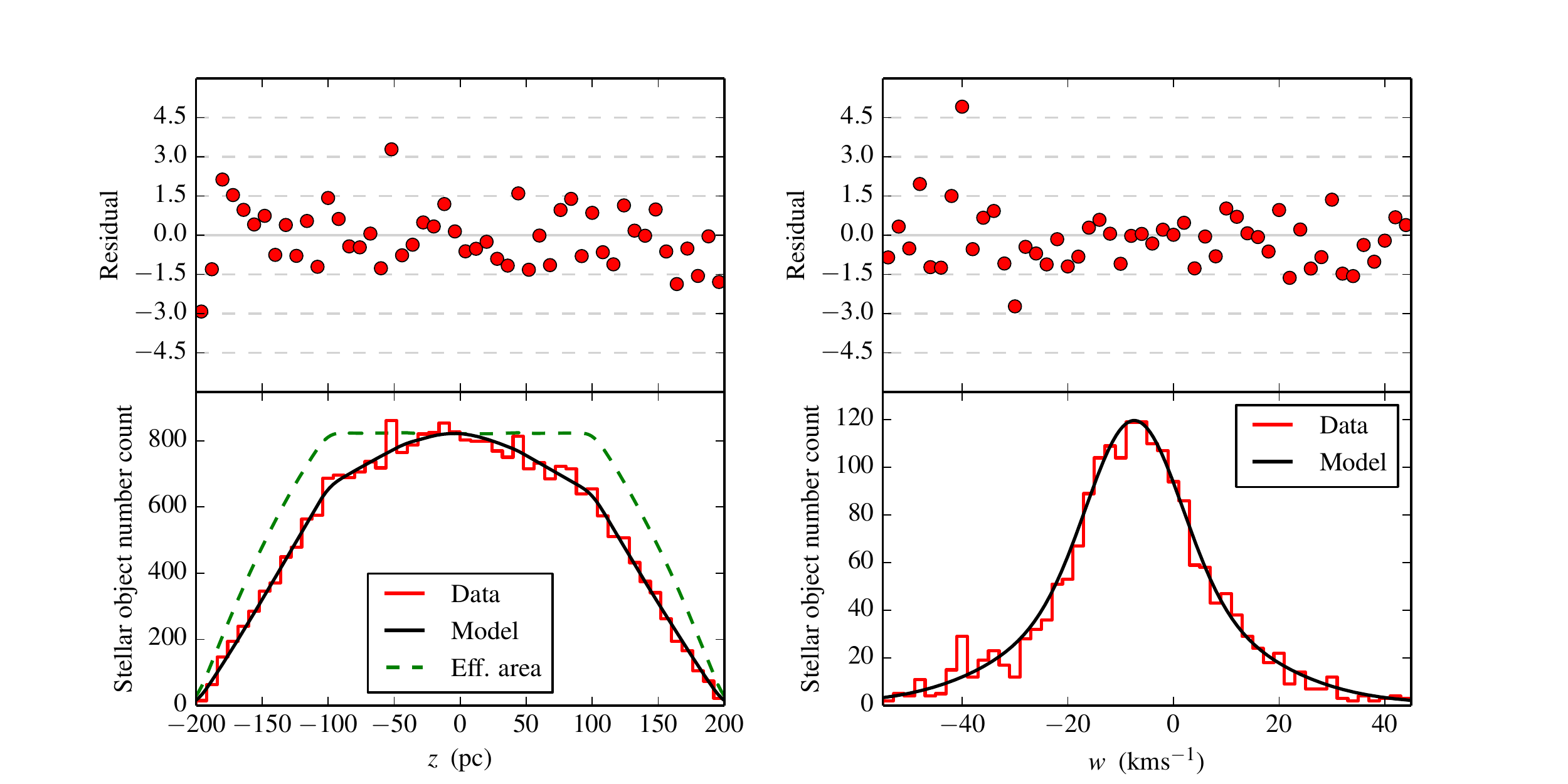}
    \caption{Control plot for sample S4, comparing the inferred model with the data. The left panel shows the stellar number count as a function of height with respect to the Sun. The right panel shows the vertical velocity distribution for stars close to the Galactic plane.}
    \label{fig:control_S4}
\end{figure*}

\begin{figure*}
	\centering
	\includegraphics[width=1.\linewidth]{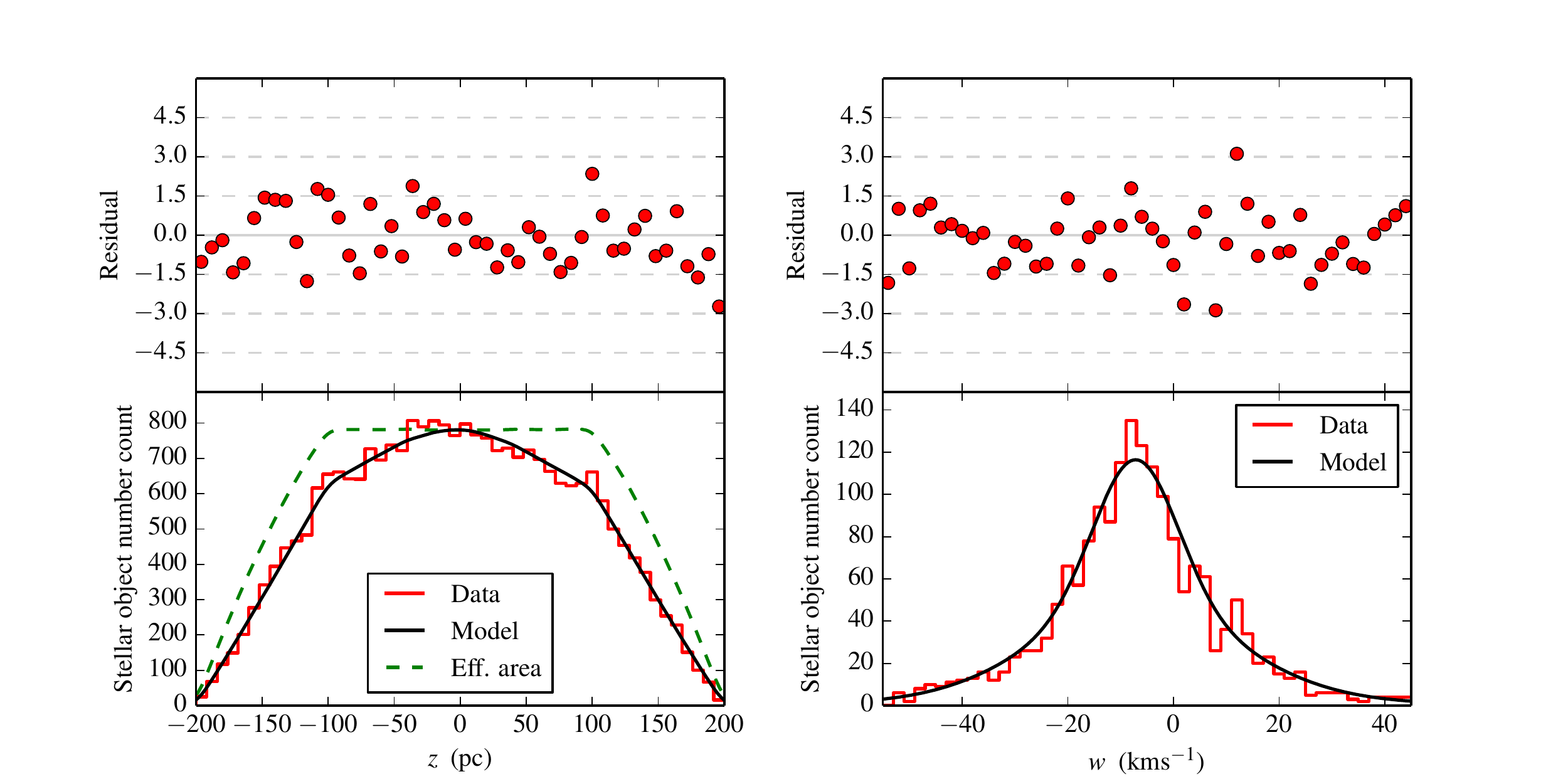}
    \caption{Control plot for sample S5, comparing the inferred model with the data. The left panel shows the stellar number count as a function of height with respect to the Sun. The right panel shows the vertical velocity distribution for stars close to the Galactic plane.}
    \label{fig:control_S5}
\end{figure*}

\begin{figure*}
	\centering
	\includegraphics[width=1.\linewidth]{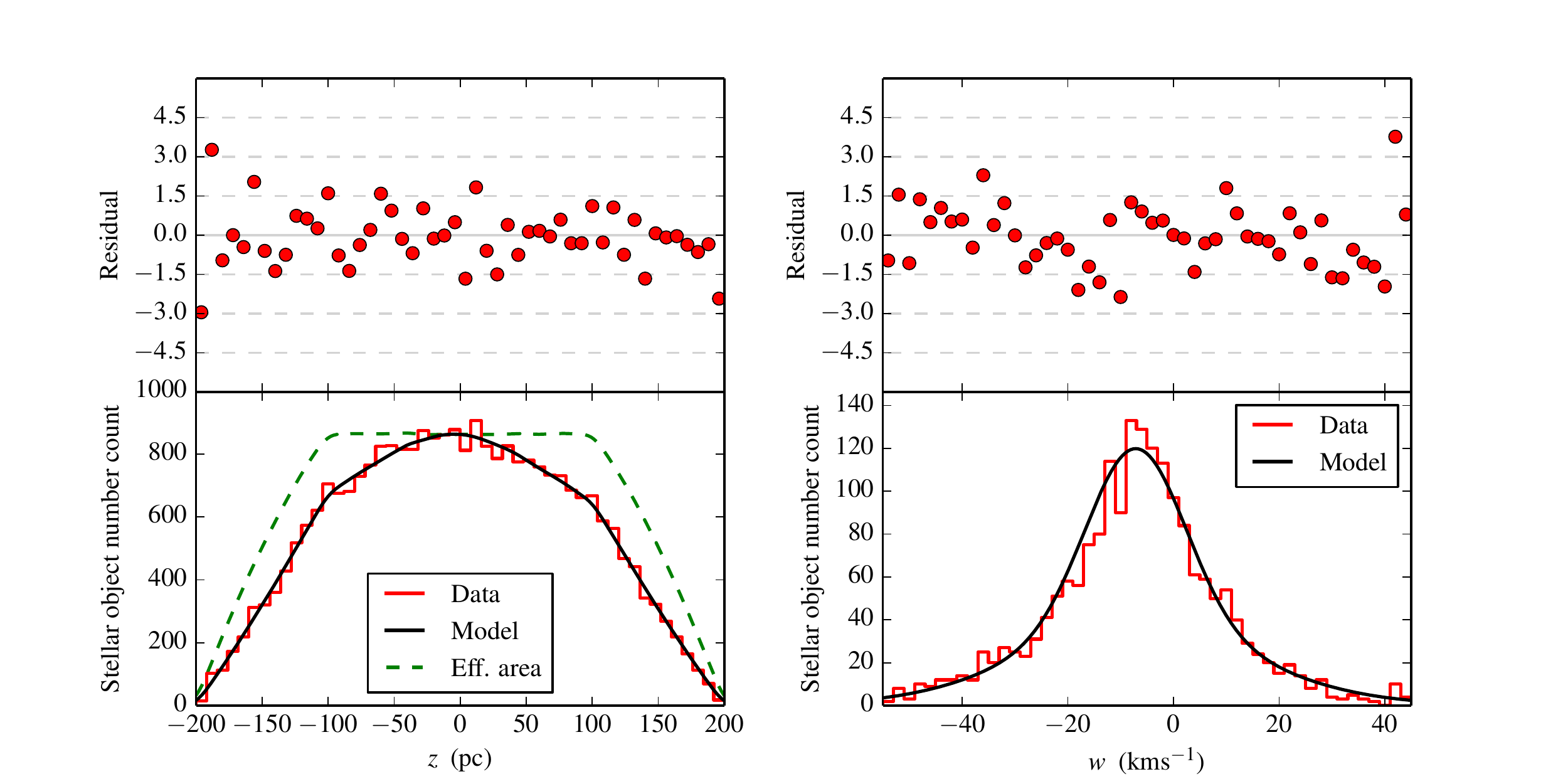}
    \caption{Control plot for sample S6, comparing the inferred model with the data. The left panel shows the stellar number count as a function of height with respect to the Sun. The right panel shows the vertical velocity distribution for stars close to the Galactic plane.}
    \label{fig:control_S6}
\end{figure*}

\begin{figure*}
	\centering
	\includegraphics[width=1.\linewidth]{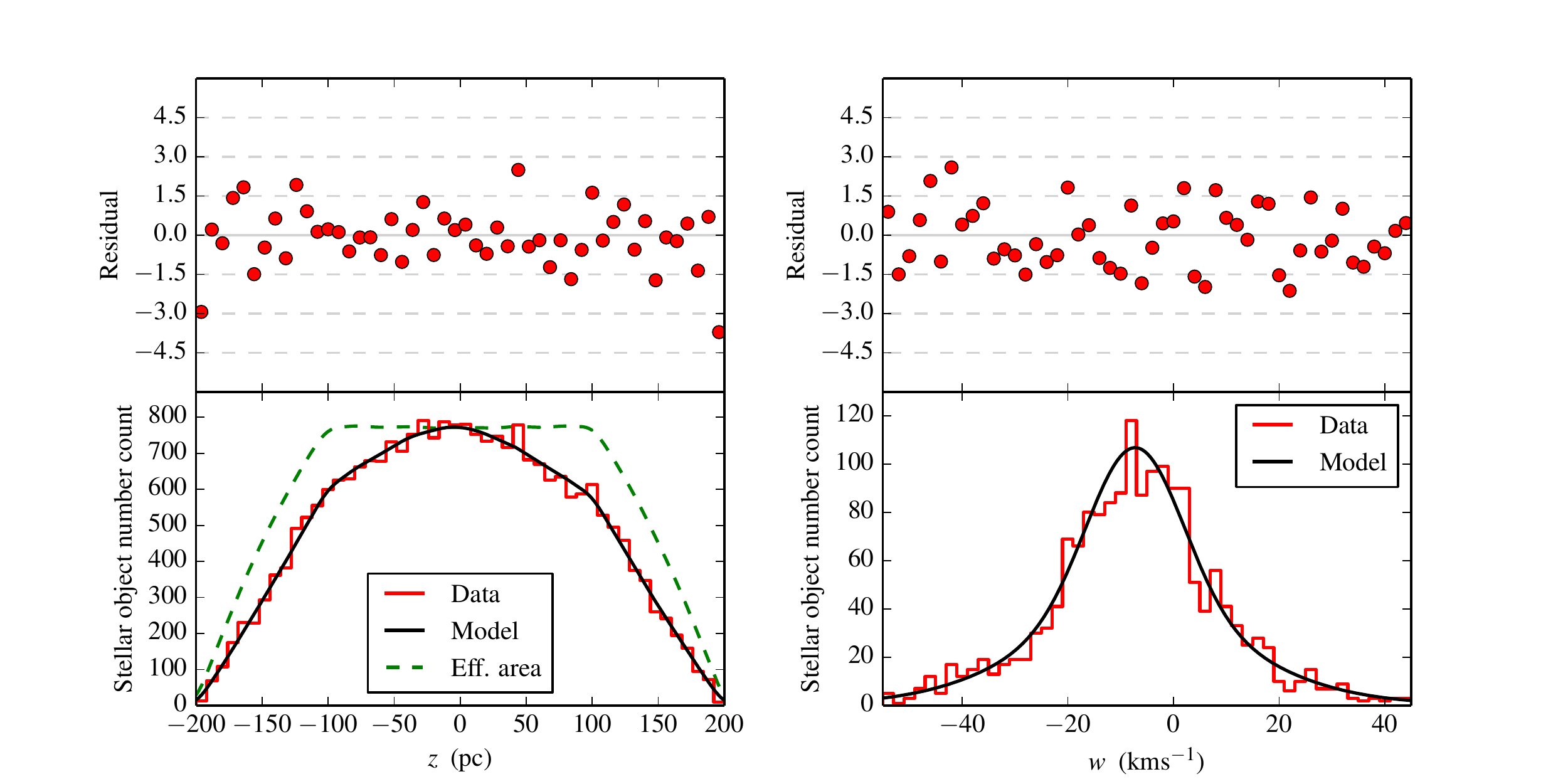}
    \caption{Control plot for sample S7, comparing the inferred model with the data. The left panel shows the stellar number count as a function of height with respect to the Sun. The right panel shows the vertical velocity distribution for stars close to the Galactic plane.}
    \label{fig:control_S7}
\end{figure*}

\begin{figure*}
	\centering
	\includegraphics[width=1.\linewidth]{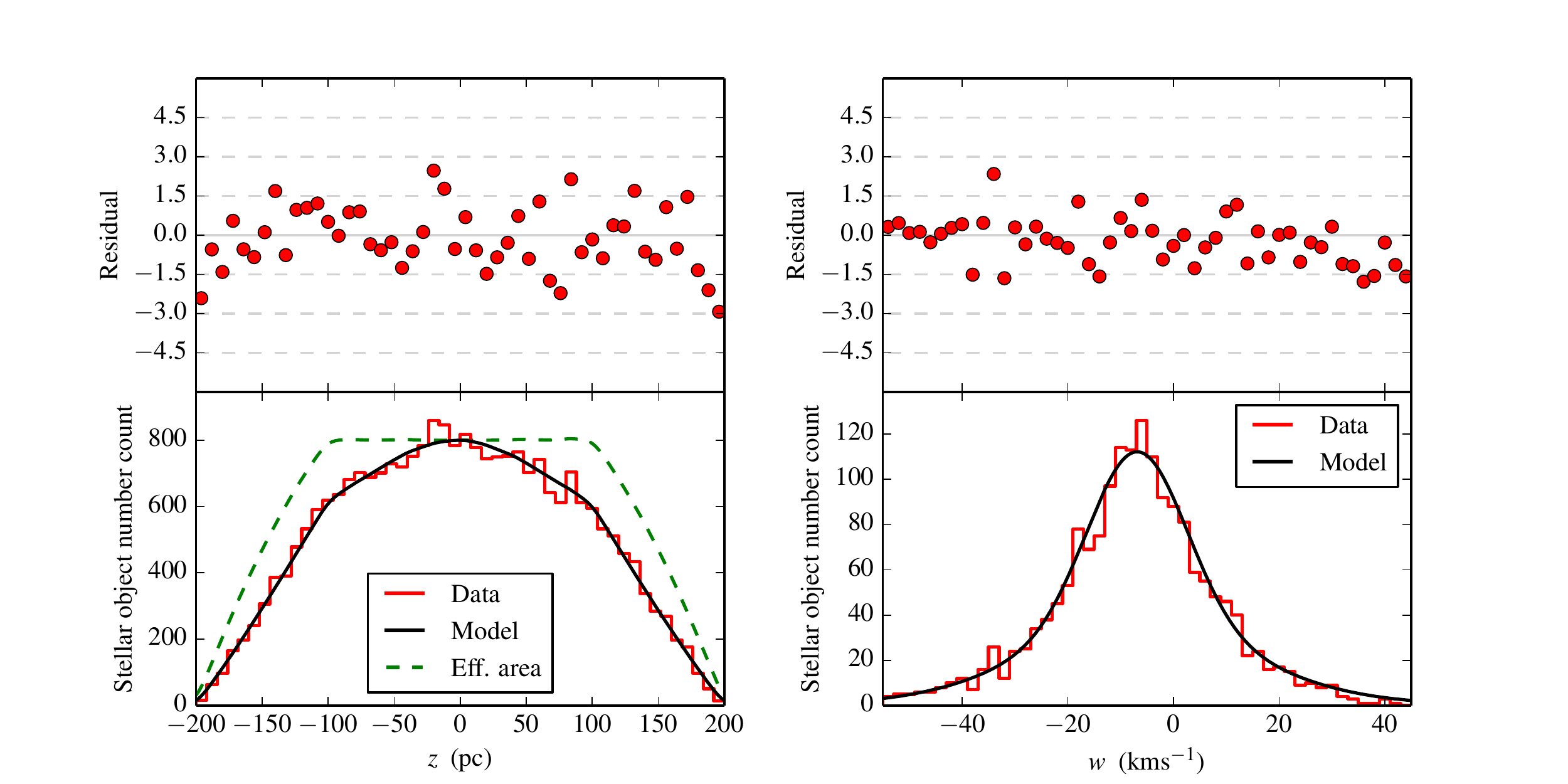}
    \caption{Control plot for sample S8, comparing the inferred model with the data. The left panel shows the stellar number count as a function of height with respect to the Sun. The right panel shows the vertical velocity distribution for stars close to the Galactic plane.}
    \label{fig:control_S8}
\end{figure*}

In Figs.~\ref{fig:control_S1}--\ref{fig:control_S8}, we show comparisons between model and data for samples S1--S8. The adopted model in these figures is given by the median values of population parameters in the inferred posterior density.

In the left panel of these figures, we show the stellar object number count in bins of height with respect to the Sun ($z$), which is equal to the stellar number density times the effective area. The effective area is the number count that would be observed if the stellar number density was uniform, owing its shape to the sample shell volume. In the figure, the effective area is normalised to be equal to the model curve in $z=0~\pc$. The data of this panel is a histogram over the full data sample, where the height is given by the angle $b$ and observed parallax $\hat{\varpi}$, thus neglecting observational errors.

In the right panel of these figures, we show the stellar object number count in bins of vertical velocity with respect to the Sun ($w$). This is for stellar objects that fulfil the following criteria: $|\hat{Z}|=|\hat{z}+Z_\odot|<10~\pc$; $\hat{\sigma}_\varpi < 0.2$ mas; either an available radial velocity measurement, or $|b|<5$ degrees. For stellar objects that have a radial velocity measurement, the velocity of that object is taken to be
\begin{equation}
    w = k_\mu \cos(b)\,\frac{\hat{\mu}_b}{\hat{\varpi}}+\sin(b)\,\hat{v}_{RV},
\end{equation}
where $k_\mu$ is the same unit conversion constant that is used in Eq.~\eqref{eq:tilde_v}. For stars without radial velocity, but $|b|<5$ degrees, the velocity is instead
\begin{equation}
    w = k_\mu \frac{\hat{\mu}_b}{\hat{\varpi}}.
\end{equation}
As such, the vertical velocity is only shown for a subset of stellar objects with small uncertainties, close to the Galactic plane. In the full statistical model, the velocity information of all stars is considered. The residuals, calculated for each separate bin, is defined as
\begin{equation}
    \text{Residual} = \frac{\text{Data}-\text{Model}}{\sqrt{\text{Model}}}.
\end{equation}


\end{appendix}

\end{document}